\documentclass[10pt]{article}
\usepackage[preprint]{tmlr}
\usepackage{amsmath}
\usepackage{amssymb}
\usepackage{booktabs}
\usepackage{graphicx}
\usepackage{hyperref}
\usepackage{url}

\newcommand{\ChanceExactNaturalSliceAucpr}{0.682366}  
\newcommand{\CicidsArmAttacks}{352962}  
\newcommand{\CicidsArmPrevalencePct}{22.060125}  
\newcommand{\CicidsArmRows}{1600000}  
\newcommand{\CicidsAttacksMovedOutOfHeldout}{103189}  
\newcommand{\CicidsEcodScoringBatch}{480000}  
\newcommand{\CicidsFridayHeldoutDensityPct}{77.66}  
\newcommand{\CicidsHeldoutAttackFreeRows}{162000}  
\newcommand{\CicidsHeldoutOverlap}{78000}  
\newcommand{\CicidsHeldoutOverlapPct}{32.5}  
\newcommand{\CicidsHeldoutSize}{240000}  
\newcommand{\CicidsNaturalHeldoutSpanMinutes}{204.2}  
\newcommand{\CicidsSyntheticHeldoutAttackDays}{1}  
\newcommand{\CicidsWeekEligibleRows}{2087997}  
\newcommand{\CicidsWeekMaxDayRetentionPct}{93.156}  
\newcommand{\CicidsWeekMinDayRetentionPct}{64.399}  
\newcommand{\CicidsWeekPrevalenceBiasPp}{-2.1464}  
\newcommand{\CicidsWeekRetentionPct}{76.628}  
\newcommand{\CicidsWeekTruePrevalencePct}{24.2065}  
\newcommand{\CicidsWeekUsedPrevalencePct}{22.0601}  
\newcommand{\CicidsWeekUsedRows}{1600000}  
\newcommand{\ContrastCicidsDim}{84}  
\newcommand{\ContrastLitnetBlasterWormDim}{36}  
\newcommand{\ContrastLitnetSpamDim}{36}  
\newcommand{\ContrastLitnetUdpFloodDim}{36}  
\newcommand{\LitnetCaptureEcodScoringBatch}{150000}  
\newcommand{\LitnetPooledEcodScoringBatch}{450000}  
\newcommand{\RefBootBTwoFiveZeroConclusionChanges}{0}  
\newcommand{\RefBootBTwoSixZeroZeroBranchCombinedAucprHi}{0.807028}  
\newcommand{\RefBootBTwoSixZeroZeroBranchCombinedAucprLo}{0.653543}  
\newcommand{\RefBootBTwoSixZeroZeroBranchTailOnlyAucprHi}{0.880632}  
\newcommand{\RefBootBTwoSixZeroZeroBranchTailOnlyAucprLo}{0.781947}  
\newcommand{\RefBootBTwoSixZeroZeroConclusionChanges}{1}  
\newcommand{\RefBootBlockLength}{100}  
\newcommand{\RefBootBranchAuxOnlyAucprHi}{0.619628}  
\newcommand{\RefBootBranchAuxOnlyAucprLo}{0.583380}  
\newcommand{\RefBootBranchAuxOnlyAucrocHi}{0.291531}  
\newcommand{\RefBootBranchAuxOnlyAucrocLo}{0.272288}  
\newcommand{\RefBootBranchCombinedAucprHi}{0.744732}  
\newcommand{\RefBootBranchCombinedAucprLo}{0.714449}  
\newcommand{\RefBootBranchCombinedAucrocHi}{0.542664}  
\newcommand{\RefBootBranchCombinedAucrocLo}{0.510641}  
\newcommand{\RefBootBranchTailOnlyAucprHi}{0.842969}  
\newcommand{\RefBootBranchTailOnlyAucprLo}{0.822025}  
\newcommand{\RefBootBranchTailOnlyAucrocHi}{0.834607}  
\newcommand{\RefBootBranchTailOnlyAucrocLo}{0.824120}  
\newcommand{\RefBootFullNaturalMarginHi}{0.032692}  
\newcommand{\RefBootFullNaturalMarginLo}{0.021099}  
\newcommand{\RefBootFullSyntheticMarginHi}{0.130294}  
\newcommand{\RefBootFullSyntheticMarginLo}{0.122046}  
\newcommand{\RefBootResamples}{1000}  
\newcommand{\RefBootRobustBlockLengthA}{250}  
\newcommand{\RefBootRobustBlockLengthB}{2600}  
\newcommand{\RefBootSharedTwoFourZeroNaturalMarginHi}{0.067720}  
\newcommand{\RefBootSharedTwoFourZeroNaturalMarginLo}{0.053856}  
\newcommand{\RefBootSharedTwoFourZeroSyntheticMarginHi}{0.056329}  
\newcommand{\RefBootSharedTwoFourZeroSyntheticMarginLo}{0.044634}  
\newcommand{\RefBootSharedSevenEightNaturalMarginHi}{0.064229}  
\newcommand{\RefBootSharedSevenEightNaturalMarginLo}{0.051770}  
\newcommand{\RefBootSharedSevenEightSyntheticMarginHi}{0.061923}  
\newcommand{\RefBootSharedSevenEightSyntheticMarginLo}{0.048620}  
\newcommand{\RefEcodCompBatchRecords}{240000}  
\newcommand{\RefEcodCompDeltaAucpr}{0.007552}  
\newcommand{\RefEcodCompDeltaAucroc}{0.009096}  
\newcommand{\RefEcodCompDifferingRecords}{162000}  
\newcommand{\RefEcodCompNaturalBatchAucpr}{0.844487}  
\newcommand{\RefEcodCompSharedRecords}{78000}  
\newcommand{\RefEcodCompSyntheticBatchAucpr}{0.852039}  
\newcommand{\RefEcodSharedSevenEightCrossArmSpread}{0.002788}  
\newcommand{\RefEcodSharedSevenEightNaturalAucpr}{0.847585}  
\newcommand{\RefEcodSharedSevenEightNaturalAucroc}{0.802003}  
\newcommand{\RefEcodSharedSevenEightSyntheticAucpr}{0.844797}  
\newcommand{\RefEcodSharedSevenEightSyntheticAucroc}{0.797670}  
\newcommand{\RefImputeCicidsAffectedFeatures}{2}  
\newcommand{\RefImputeCicidsAffectedRows}{4}  
\newcommand{\RefImputeCicidsInfCells}{8}  
\newcommand{\RefImputeCicidsRows}{1600000}  
\newcommand{\RefImputeLitnetBlasterWormAffectedRows}{0}  
\newcommand{\RefImputeLitnetSpamAffectedRows}{0}  
\newcommand{\RefImputeLitnetUdpFloodAffectedRows}{0}  
\newcommand{\RefMarginSharedSevenEightNatural}{0.058028}  
\newcommand{\RefMarginSharedSevenEightSynthetic}{0.055574}  
\newcommand{\RefPairedCutSixZeroNaturalDetectorAucpr}{0.483383}  
\newcommand{\RefPairedCutSixZeroNaturalEcodAucpr}{0.451502}  
\newcommand{\RefPairedCutSixZeroNaturalPrevalence}{0.350142}  
\newcommand{\RefPairedCutSixZeroOrderingsAgree}{1}  
\newcommand{\RefPairedCutSixZeroSyntheticDetectorAucpr}{0.395789}  
\newcommand{\RefPairedCutSixZeroSyntheticEcodAucpr}{0.386564}  
\newcommand{\RefPairedCutSixZeroSyntheticPrevalence}{0.254153}  
\newcommand{\RefPairedCutSixFiveNaturalDetectorAucpr}{0.528435}  
\newcommand{\RefPairedCutSixFiveNaturalEcodAucpr}{0.498989}  
\newcommand{\RefPairedCutSixFiveNaturalPrevalence}{0.400020}  
\newcommand{\RefPairedCutSixFiveOrderingsAgree}{1}  
\newcommand{\RefPairedCutSixFiveSyntheticDetectorAucpr}{0.402082}  
\newcommand{\RefPairedCutSixFiveSyntheticEcodAucpr}{0.385961}  
\newcommand{\RefPairedCutSixFiveSyntheticPrevalence}{0.240532}  
\newcommand{\RefPairedCutSevenZeroNaturalDetectorAucpr}{0.578913}  
\newcommand{\RefPairedCutSevenZeroNaturalEcodAucpr}{0.561230}  
\newcommand{\RefPairedCutSevenZeroNaturalPrevalence}{0.461644}  
\newcommand{\RefPairedCutSevenZeroOrderingsAgree}{1}  
\newcommand{\RefPairedCutSevenZeroSyntheticDetectorAucpr}{0.436118}  
\newcommand{\RefPairedCutSevenZeroSyntheticEcodAucpr}{0.389094}  
\newcommand{\RefPairedCutSevenZeroSyntheticPrevalence}{0.234708}  
\newcommand{\RefPairedCutSevenFiveNaturalDetectorAucpr}{0.508857}  
\newcommand{\RefPairedCutSevenFiveNaturalEcodAucpr}{0.556352}  
\newcommand{\RefPairedCutSevenFiveNaturalPrevalence}{0.429028}  
\newcommand{\RefPairedCutSevenFiveOrderingsAgree}{0}  
\newcommand{\RefPairedCutSevenFiveSyntheticDetectorAucpr}{0.477333}  
\newcommand{\RefPairedCutSevenFiveSyntheticEcodAucpr}{0.389087}  
\newcommand{\RefPairedCutSevenFiveSyntheticPrevalence}{0.229687}  
\newcommand{\RefPairedCutEightZeroNaturalDetectorAucpr}{0.585598}  
\newcommand{\RefPairedCutEightZeroNaturalEcodAucpr}{0.639969}  
\newcommand{\RefPairedCutEightZeroNaturalPrevalence}{0.513066}  
\newcommand{\RefPairedCutEightZeroOrderingsAgree}{0}  
\newcommand{\RefPairedCutEightZeroSyntheticDetectorAucpr}{0.536930}  
\newcommand{\RefPairedCutEightZeroSyntheticEcodAucpr}{0.400022}  
\newcommand{\RefPairedCutEightZeroSyntheticPrevalence}{0.241975}  
\newcommand{\RefPairedCutEightFiveNaturalDetectorAucpr}{0.728355}  
\newcommand{\RefPairedCutEightFiveNaturalEcodAucpr}{0.758205}  
\newcommand{\RefPairedCutEightFiveNaturalPrevalence}{0.682350}  
\newcommand{\RefPairedCutEightFiveOrderingsAgree}{0}  
\newcommand{\RefPairedCutEightFiveSyntheticDetectorAucpr}{0.544998}  
\newcommand{\RefPairedCutEightFiveSyntheticEcodAucpr}{0.417522}  
\newcommand{\RefPairedCutEightFiveSyntheticPrevalence}{0.252396}  
\newcommand{\RefPairedCutNineZeroNaturalDetectorAucpr}{0.799279}  
\newcommand{\RefPairedCutNineZeroNaturalEcodAucpr}{0.791671}  
\newcommand{\RefPairedCutNineZeroNaturalPrevalence}{0.716250}  
\newcommand{\RefPairedCutNineZeroOrderingsAgree}{1}  
\newcommand{\RefPairedCutNineZeroSyntheticDetectorAucpr}{0.574244}  
\newcommand{\RefPairedCutNineZeroSyntheticEcodAucpr}{0.474453}  
\newcommand{\RefPairedCutNineZeroSyntheticPrevalence}{0.285544}  
\newcommand{\RefPairedCuts}{7}  
\newcommand{\RefPairedCutsOrderingsAgree}{4}  
\newcommand{\RefPairedCutsOrderingsDiffer}{3}  
\newcommand{\RefRelocatedToTraining}{77670}  
\newcommand{\RefRelocatedToValidation}{25519}  
\newcommand{\RevBranchAuxBindsHeldoutPct}{58.452}  
\newcommand{\RevBranchAuxOnlyAucpr}{0.600270}  
\newcommand{\RevBranchAuxOnlyAucroc}{0.281890}  
\newcommand{\RevBranchCombinedAucpr}{0.728355}  
\newcommand{\RevBranchCombinedAucroc}{0.526623}  
\newcommand{\RevBranchTailMinusCombinedAucpr}{0.103477}  
\newcommand{\RevBranchTailMinusCombinedAucroc}{0.302658}  
\newcommand{\RevBranchTailOnlyAucpr}{0.831832}  
\newcommand{\RevBranchTailOnlyAucroc}{0.829281}  
\newcommand{\RevEcodBatchTwoFourZeroZeroZeroZeroAucpr}{0.758205}  
\newcommand{\RevEcodBatchThreeZeroZeroZeroZeroZeroAucpr}{0.760029}  
\newcommand{\RevEcodBatchThreeSixZeroZeroZeroZeroAucpr}{0.762108}  
\newcommand{\RevEcodBatchFourEightZeroZeroZeroZeroAucpr}{0.755142}  
\newcommand{\RevEcodBatchDelta}{0.003063}  
\newcommand{\RevEcodBatchEvalRecords}{240000}  
\newcommand{\RevEcodBatchLadderOne}{300000}  
\newcommand{\RevEcodBatchLadderTwo}{360000}  
\newcommand{\RevEcodBatchSpread}{0.006966}  
\newcommand{\RevReproNaturalDelta}{0.000018}  
\newcommand{\RevSharedAttacks}{60575}  
\newcommand{\RevSharedDetectorNaturalAucpr}{0.905613}  
\newcommand{\RevSharedDetectorNaturalAucroc}{0.869219}  
\newcommand{\RevSharedDetectorSpread}{0.005242}  
\newcommand{\RevSharedDetectorSyntheticAucpr}{0.900371}  
\newcommand{\RevSharedDetectorSyntheticAucroc}{0.864159}  
\newcommand{\RevSharedEcodNaturalAucpr}{0.844487}  
\newcommand{\RevSharedEcodNaturalAucroc}{0.799910}  
\newcommand{\RevSharedEcodSyntheticAucpr}{0.849842}  
\newcommand{\RevSharedEcodSyntheticAucroc}{0.805657}  
\newcommand{\RevSharedMarginNatural}{0.061126}  
\newcommand{\RevSharedMarginSynthetic}{0.050529}  
\newcommand{\RevSharedPrevalence}{0.776603}  
\newcommand{\RevSharedRecords}{78000}  
\newcommand{\RevSplitCuts}{7}  
\newcommand{\RevSplitEcodWins}{3}  
\newcommand{\SFourCicidsInterleavedTestAttacks}{60575}  
\newcommand{\SFourCicidsInterleavedTestPrev}{25.2396}  
\newcommand{\SFourCicidsNaturalChanceFloor}{0.682350}  
\newcommand{\SFourCicidsNaturalEcodAucpr}{0.755142}  
\newcommand{\SFourCicidsNaturalEcodMinusDetectorAucpr}{0.026805}  
\newcommand{\SFourCicidsNaturalEcodNormLift}{0.229158}  
\newcommand{\SFourCicidsNaturalEcodOwnBatchMinusDetectorAucpr}{0.029868}  
\newcommand{\SFourCicidsNaturalHstAucpr}{0.588902}  
\newcommand{\SFourCicidsNaturalHstLiftAbs}{0.093448}  
\newcommand{\SFourCicidsNaturalProposedDetectorAucpr}{0.728337}  
\newcommand{\SFourCicidsNaturalProposedDetectorNormLift}{0.144773}  
\newcommand{\SFourCicidsNaturalTestAttacks}{163764}  
\newcommand{\SFourCicidsNaturalTestPrev}{68.235}  
\newcommand{\SFourCicidsPrevShiftAbsPp}{42.9954}  
\newcommand{\SFourCicidsSyntheticChanceFloor}{0.252396}  
\newcommand{\SFourCicidsSyntheticDetectorMinusEcodAucpr}{0.126032}  
\newcommand{\SFourCicidsSyntheticEcodAucpr}{0.418966}  
\newcommand{\SFourCicidsSyntheticEcodNormLift}{0.222805}  
\newcommand{\SFourCicidsSyntheticProposedDetectorAucpr}{0.544998}  
\newcommand{\SFourCicidsSyntheticProposedDetectorNormLift}{0.391386}  
\newcommand{\SFourLitnetBlasterWormEcodAucpr}{0.035710}  
\newcommand{\SFourLitnetBlasterWormHstAucpr}{0.147731}  
\newcommand{\SFourLitnetBlasterWormProposedDetectorAucpr}{0.964091}  
\newcommand{\SFourLitnetBlasterWormTestAttacks}{2658}  
\newcommand{\SFourLitnetBlasterWormTestPrev}{3.544}  
\newcommand{\SFourLitnetPooledEcodAucpr}{0.229143}  
\newcommand{\SFourLitnetPooledHstAucpr}{0.238840}  
\newcommand{\SFourLitnetPooledProposedDetectorAucpr}{0.943056}  
\newcommand{\SFourLitnetPooledTestAttacks}{14621}  
\newcommand{\SFourLitnetPooledTestPrev}{6.4982}  
\newcommand{\SFourLitnetSpamEcodAucpr}{0.106805}  
\newcommand{\SFourLitnetSpamHstAucpr}{0.008447}  
\newcommand{\SFourLitnetSpamProposedDetectorAucpr}{0.846154}  
\newcommand{\SFourLitnetSpamTestAttacks}{132}  
\newcommand{\SFourLitnetSpamTestPrev}{0.176}  
\newcommand{\SFourLitnetUdpFloodEcodAucpr}{0.321185}  
\newcommand{\SFourLitnetUdpFloodHstAucpr}{0.552675}  
\newcommand{\SFourLitnetUdpFloodProposedDetectorAucpr}{0.394428}  
\newcommand{\SFourLitnetUdpFloodTestAttacks}{11831}  
\newcommand{\SFourLitnetUdpFloodTestPrev}{15.7747}  
\newcommand{\SFourSeedCicidsTwoZeroOneSevenInterleavedSyntheticHstDrawOne}{0.513623}  
\newcommand{\SFourSeedCicidsTwoZeroOneSevenInterleavedSyntheticHstDrawThree}{0.358504}  
\newcommand{\SFourSeedCicidsTwoZeroOneSevenInterleavedSyntheticHstDrawTwo}{0.427004}  
\newcommand{\SFourSeedCicidsTwoZeroOneSevenInterleavedSyntheticHstMean}{0.433044}  
\newcommand{\SFourSeedCicidsTwoZeroOneSevenInterleavedSyntheticHstSd}{0.077736}  
\newcommand{\SFourSeedLitnetPooledCompositeSyntheticHstDrawOne}{0.238840}  
\newcommand{\SFourSeedLitnetPooledCompositeSyntheticHstDrawThree}{0.367761}  
\newcommand{\SFourSeedLitnetPooledCompositeSyntheticHstDrawTwo}{0.177599}  
\newcommand{\SSixChanceFloor}{0.161333}  
\newcommand{\SSixCorrectedAucpr}{0.169912}  
\newcommand{\SSixCorrectedAucroc}{0.509620}  
\newcommand{\SSixCorrectedNormLift}{0.010229}  
\newcommand{\SSixDiagCorrectedDistinctScores}{747}  
\newcommand{\SSixDiagCorrectedFracAtCap}{0.9274}  
\newcommand{\SSixDiagCorrectedMeanShortRunMass}{1}  
\newcommand{\SSixDiagCorrectedScoreSd}{0.159341}  
\newcommand{\SSixDiagOriginalDistinctScores}{3899}  
\newcommand{\SSixDiagOriginalFracAtCap}{0.004733}  
\newcommand{\SSixDiagOriginalMeanShortRunMass}{0.006472}  
\newcommand{\SSixDiagOriginalScoreSd}{0.218999}  
\newcommand{\SSixDiagWindowRows}{15000}  
\newcommand{\SSixOriginalAucpr}{0.397486}  
\newcommand{\SSixOriginalAucroc}{0.848216}  
\newcommand{\SSixOriginalNormLift}{0.281581}  
\newcommand{\SSixPrefixRows}{200000}  
\newcommand{\SSixProbeCorrectedPeak}{1}  
\newcommand{\SSixProbeOriginalPeak}{0.001}  
\newcommand{\SSixTestAttacks}{4840}  
\newcommand{\SSixTestRows}{30000}  
\newcommand{\SThreeBindingRows}{49970}  
\newcommand{\SThreeChangePointBindingMeanValue}{0.002531}  
\newcommand{\SThreeChangePointBindingPct}{75.039}  
\newcommand{\SThreeLatencyUsesAClock}{0}  
\newcommand{\SThreePrZeroPreMean}{0.001}  
\newcommand{\SThreePrZeroResponsePeak}{0.001}  
\newcommand{\SThreePrZeroSteadyMaxDeviation}{0}  
\newcommand{\SThreePrZeroTruncatedMax}{1}  
\newcommand{\SThreePrZeroTruncatedMean}{0.01099}  
\newcommand{\SThreeProbeHazard}{0.001}  
\newcommand{\SThreeRunLengthCap}{500}  
\newcommand{\SThreeThresholdCicids}{0.261745}  
\newcommand{\SThreeThresholdCostOnly}{0.090909}  
\newcommand{\SThreeThresholdLitnet}{0.655172}  
\newcommand{\STwoLFiveEcodMean}{0.166239}  
\newcommand{\STwoLFiveFloor}{0.049524}  
\newcommand{\STwoLFiveHstMean}{0.199261}  
\newcommand{\STwoLFiveLodaMean}{0.207581}  
\newcommand{\STwoLFiveLofMean}{0.537596}  
\newcommand{\STwoLFiveProposedLift}{0.317640}  
\newcommand{\STwoLFiveProposedMean}{0.367164}  
\newcommand{\STwoLFiveProposedNormLift}{0.334190}  
\newcommand{\STwoLFortyEcodMean}{0.527364}  
\newcommand{\STwoLFortyFloor}{0.399995}  
\newcommand{\STwoLFortyHstMean}{0.476031}  
\newcommand{\STwoLFortyLodaMean}{0.399422}  
\newcommand{\STwoLFortyLofMean}{0.896968}  
\newcommand{\STwoLFortyProposedLift}{0.094222}  
\newcommand{\STwoLFortyProposedMean}{0.494217}  
\newcommand{\STwoLFortyProposedNormLift}{0.157035}  
\newcommand{\STwoLSixtyFourEcodMean}{0.678404}  
\newcommand{\STwoLSixtyFourFloor}{0.639991}  
\newcommand{\STwoLSixtyFourHstMean}{0.499341}  
\newcommand{\STwoLSixtyFourLodaMean}{0.524691}  
\newcommand{\STwoLSixtyFourLofMean}{0.904855}  
\newcommand{\STwoLSixtyFourProposedLift}{-0.020445}  
\newcommand{\STwoLSixtyFourProposedMean}{0.619546}  
\newcommand{\STwoLSixtyFourProposedNormLift}{-0.056790}  
\newcommand{\STwoLTenEcodMean}{0.252198}  
\newcommand{\STwoLTenFloor}{0.099004}  
\newcommand{\STwoLTenHstMean}{0.299702}  
\newcommand{\STwoLTenLodaMean}{0.247426}  
\newcommand{\STwoLTenLofMean}{0.697003}  
\newcommand{\STwoLTenProposedLift}{0.348209}  
\newcommand{\STwoLTenProposedMean}{0.447213}  
\newcommand{\STwoLTenProposedNormLift}{0.386471}  
\newcommand{\STwoLUnresampledEcodMean}{0.418966}  
\newcommand{\STwoLUnresampledFloor}{0.252396}  
\newcommand{\STwoLUnresampledHstMean}{0.433044}  
\newcommand{\STwoLUnresampledLodaMean}{0.341715}  
\newcommand{\STwoLUnresampledLofMean}{0.863194}  
\newcommand{\STwoLUnresampledProposedLift}{0.292602}  
\newcommand{\STwoLUnresampledProposedMean}{0.544998}  
\newcommand{\STwoLUnresampledProposedNormLift}{0.391386}  
\newcommand{\StreamCicidsTwoZeroOneSevenPrevalencePct}{22.0601}  
\newcommand{\StreamCicidsTwoZeroOneSevenRows}{1600000}  
\newcommand{\StreamCicidsTwoZeroOneSevenRunMax}{2522}  
\newcommand{\StreamCicidsTwoZeroOneSevenRunMedian}{2}  
\newcommand{\StreamCicidsTwoZeroOneSevenRunPninety}{70}  
\newcommand{\StreamCicidsTwoZeroOneSevenSpanMinutes}{6246.71}  
\newcommand{\StreamCicidsTwoZeroOneSevenTestPrevalencePct}{68.235}  
\newcommand{\StreamLitnetBlasterWormPrevalencePct}{0.6562}  
\newcommand{\StreamLitnetBlasterWormRows}{500000}  
\newcommand{\StreamLitnetBlasterWormRunMax}{5}  
\newcommand{\StreamLitnetBlasterWormRunMedian}{1}  
\newcommand{\StreamLitnetBlasterWormRunPninety}{2}  
\newcommand{\StreamLitnetBlasterWormSpanMinutes}{1.62}  
\newcommand{\StreamLitnetBlasterWormTestPrevalencePct}{3.544}  
\newcommand{\StreamLitnetSpamPrevalencePct}{0.0276}  
\newcommand{\StreamLitnetSpamRows}{500000}  
\newcommand{\StreamLitnetSpamRunMax}{2}  
\newcommand{\StreamLitnetSpamRunMedian}{1}  
\newcommand{\StreamLitnetSpamRunPninety}{1}  
\newcommand{\StreamLitnetSpamSpanMinutes}{39122.23}  
\newcommand{\StreamLitnetSpamTestPrevalencePct}{0.176}  
\newcommand{\StreamLitnetUdpFloodPrevalencePct}{14.9384}  
\newcommand{\StreamLitnetUdpFloodRows}{500000}  
\newcommand{\StreamLitnetUdpFloodRunMax}{20}  
\newcommand{\StreamLitnetUdpFloodRunMedian}{1}  
\newcommand{\StreamLitnetUdpFloodRunPninety}{2}  
\newcommand{\StreamLitnetUdpFloodSpanMinutes}{3.97}  
\newcommand{\StreamLitnetUdpFloodTestPrevalencePct}{15.7747}  
\newcommand{\SweepEcodScoringBatchMax}{480000}  
\newcommand{\SweepEcodScoringBatchMin}{160917}  

\title{Stream Assembly Is an Uncontrolled Treatment in\\
Streaming Intrusion-Detection Benchmarks}
\hypersetup{pdftitle={Stream Assembly Is an Uncontrolled Treatment in Streaming Intrusion-Detection Benchmarks},
            pdfauthor={}, pdfsubject={}, pdfkeywords={}}

\author{\name Michel Youssef \email michelyoussef@hotmail.com \\
      \addr Independent Researcher, Beirut, Lebanon \\
      ORCID 0009-0000-0664-8228}

\begin{document}

\maketitle

\begin{abstract}
Public network captures are rarely usable as evaluation streams as they stand,
so streaming intrusion-detection studies assemble them: interleaving capture
days, pooling temporally disjoint captures, or replaying records round robin.
We show on two benchmarks that this assembly step is not neutral plumbing but
an \emph{uncontrolled experimental treatment}, a benchmark-construction choice
that changes what the evaluation measures. On CICIDS2017, holding the full
record multiset identical and changing only the ordering, a fixed positional
70/15/15 split then produces held-out samples that share only
\CicidsHeldoutOverlapPct\%\ of their records, at held-out prevalences of
\SFourCicidsNaturalTestPrev\%\ and \SFourCicidsInterleavedTestPrev\%\ --- a
\SFourCicidsPrevShiftAbsPp-point difference --- and the measured ordering of
the two deterministic scorers reverses. Restricting both arms to the
\RevSharedRecords{} records they both held out removes that reversal: the same
scorer leads in both arms there. The reversal is therefore attributable to
which records the assembly hands to the test set, not to the order in which
the detector saw its history. That attribution assumes that history contributes
no more on the records the arms do not share than on those they do.
On LITNET-2020, pooling three temporally disjoint captures reports a single
\SFourLitnetPooledTestPrev\%\ operating point that is the equal-weight mean of
per-capture held-out prevalences spanning \SFourLitnetSpamTestPrev\%\ to
\SFourLitnetUdpFloodTestPrev\%; we present that identity as an audit check
rather than a discovery. We also audit the evaluated detector against its
description: its reset and growth branches share a predictive term that
cancels, so the reset posterior $P(r_t{=}0)$ equals the hazard rate exactly
below the run-length cap, while the evaluations spend nearly all their length at or
beyond that cap, where the posterior instead wanders; and the score the
evaluation consumes is a function of $P(r\le5)$, not of $P(r{=}0)$. Scoring
that detector one branch at a time yields a separate result: its deployed
$\max$ composition ranks \emph{worse than its own tail term alone}
(\RevBranchTailMinusCombinedAucpr{} $AP$, \RevBranchTailMinusCombinedAucroc{}
AUC-ROC), because the auxiliary branch is inverted rather than uninformative
(AUC-ROC \RevBranchAuxOnlyAucroc) and a maximum lets it set the record's score
wherever the tail is small --- a defect no metric computed on the assembled
score can attribute. Finally, we quantify a batch dependence in the ECOD
reference implementation, whose empirical CDFs are recomputed over the
training matrix concatenated with the scored batch: holding the evaluated
records and the fitted model fixed and changing only the accompanying batch
moves its AUC-PR by \RevEcodBatchDelta, so published ECOD numbers are not
comparable across studies that score a different batch, in size or composition. Every measured
value traces to an archived, hash-verified run manifest, and the
sentence-level claim ledger ships with the artifact.
\end{abstract}

\section{Introduction}
\label{sec:intro}

Machine-learning evaluations for network intrusion detection rest on a small
number of public benchmarks, and a substantial literature questions what those
benchmarks measure~\citep{catillo2023hype,arp2022dosdonts,ring2019datasets}.
This paper contributes a quantified case study and mechanism audit of one step
in that pipeline: the step that assembles capture files into an evaluation
stream. CICIDS2017 spans five capture days of very different attack
density~\citep{sharafaldin2018cicids,engelen2021cicids}, while LITNET-2020
annotates twelve attack types collected over about ten
months~\citep{damasevicius2020litnet} and the three captures we evaluate are
temporally disjoint, so evaluations
interleave, pool, or replay records to obtain a single stream with workable
splits. We measure what that step does on these two benchmarks.

\subsection*{What this paper claims, and what it does not}
Our CICIDS2017 intervention changes the ordering of a fixed record multiset and
holds the split \emph{rule} fixed. It does not hold the evaluated \emph{sample}
fixed: under a positional split, reordering changes which records land in
training, validation and test, and therefore changes held-out membership,
held-out prevalence, and within-split order simultaneously. Stream assembly is
in this precise sense an \textbf{uncontrolled treatment}. We therefore make a
pipeline-level claim --- assembling the stream differently changes the measured
evaluation outcome --- and we do \emph{not} claim that order alone causes the
change, nor that attack prevalence is excluded as an explanation. Isolating
those factors requires a factorial or prevalence-matched design that we
identify as the natural next experiment.

Two further scoping statements belong here rather than in a limitations
section. First, we make no claim about how often the practice occurs: we have
not conducted a systematic literature survey, and the construction we contrast
against is the one used by the audited earlier evaluation of this same work,
with interleaving and pooling also discussed in the cited benchmark-criticism
literature. Second, the conceptual warning that arbitrary temporal ordering and
sampling can distort security-ML evaluation is established prior
art~\citep{catillo2023hype,arp2022dosdonts}; our contribution is the exact
same-multiset measurement on a real benchmark, with overlap, relocation,
prevalence and outcome reported together, plus the audit of the released
detector.

\subsection*{Origin of this version}
Earlier versions of this manuscript (arXiv:2605.24696, v1 and v2) reported results produced under the pooled
and interleaved constructions studied here, and described a scoring rule the
released code did not implement. An independent adversarial review and a
line-by-line audit of the archived artifacts established both. The present
version is the rebuild from that audit: the assembled constructions are
retained as explicitly labelled protocols and contrasted against
least-transformed alternatives; the detector is described as what the code
computes; and every number was regenerated --- or, for the retained prevalence
sweep, re-derived from the archived pre-audit runs --- by a script that wrote,
in the same execution, an archived manifest recording the git commit,
configuration hash, input SHA-256, seed, environment hash and output path.
Section~\ref{sec:disclosure} states which measurements this paper shares with
those versions and which are new, and gives the per-result-group account; the
dated correction history is the corrected-incident log in the artifact.

\subsection*{Where this work sits}
Read as machine-learning evaluation methodology, this paper makes three points
and uses the intrusion-detection corpora as the instance. First, benchmark
construction is an experimental treatment: the choices that turn raw captures
into an evaluation stream change the held-out sample, its class prevalence and
its order at once, and so change what a reported number measures. The
benchmark-criticism literature describes effects of this kind for temporal
bias~\citep{pendlebury2019tesseract}, for the evaluation of anomaly
detectors~\citep{campos2016evaluation,wu2023flawed}, for how much a
network-intrusion result depends on which corpora are
combined~\citep{apruzzese2022crosseval}, and for security machine learning
generally~\citep{arp2022dosdonts,catillo2023hype}; this paper measures one
such effect on a fixed multiset. Second, a widely used library scores
transductively: ECOD in PyOD recomputes its empirical distributions over the
training matrix concatenated with the batch it is asked to score, so a fixed
record's score depends on the batch it is scored within, and we quantify that
dependence with the evaluated records and the fitted model held fixed, across
batch size and, separately, at fixed batch size. Third, reporting can be made
provenance-enforced: every measured value here resolves through a generated
macro to an archived run manifest, and the build fails on any that does not,
which addresses the reproducibility gap documented for this
field~\citep{olszewski2023reproducibility}.

\subsection*{Contributions}
\begin{enumerate}
\item \textbf{Stream assembly as an uncontrolled treatment}
(Section~\ref{sec:contrast}). On the identical \CicidsArmRows-record multiset
(\CicidsArmAttacks{} attacks, whole-stream prevalence
\CicidsArmPrevalencePct\%, permutation verified mechanically), reordering under
a fixed positional split yields held-out slices sharing
\CicidsHeldoutOverlap{} of \CicidsHeldoutSize{} records
(\CicidsHeldoutOverlapPct\%), moves held-out prevalence by
\SFourCicidsPrevShiftAbsPp{} points, relocates
\CicidsAttacksMovedOutOfHeldout{} attacks out of the held-out slice, and
reverses the measured ordering of the two deterministic scorers ---
a reversal that does \emph{not} survive when both arms are restricted to the
records they both held out (Section~\ref{sec:shared}).
\item \textbf{A pooling identity, offered as an audit check}
(Section~\ref{sec:contrast-litnet}). LITNET's pooled
\SFourLitnetPooledTestPrev\%\ is the equal-weight mean of the three per-capture
held-out prevalences, forced by equal budgets and a divisible round-robin
boundary. It is arithmetic, not a discovery; its value is diagnostic --- a
pooled operating point can correspond to no capture in the pool.
\item \textbf{A scoped method-identity audit} (Section~\ref{sec:detector}).
Below the run-length cap the reset and growth branches share a predictive term
that cancels, so the reset posterior $P(r_t{=}0)$ equals the hazard rate for
any data, a property of the published recursion~\citep{adams2007bocpd} and not
a coding error; at and beyond the cap --- where the evaluated runs spend
nearly all their length --- truncation breaks that identity and the posterior
wanders. The score the evaluation consumes is a function of $P(r_t\le5)$, not
the $P(r_t{=}0)$ the earlier description defined it as. We state these three
facts separately because only the first is an exact result.
\item \textbf{A composition defect that costs more than it appears to}
(Section~\ref{sec:branch}). Scored branch-wise on the timestamp-ordered
held-out slice, the deployed $\max$ composition ranks \emph{worse than its own
tail term alone} --- by \RevBranchTailMinusCombinedAucpr{} $AP$ and
\RevBranchTailMinusCombinedAucroc{} AUC-ROC. The auxiliary branch is not
uninformative but inverted (AUC-ROC \RevBranchAuxOnlyAucroc), and because the
score is a maximum it sets the record's score wherever the tail is small,
dragging the deployed detector to near-chance AUC-ROC (\RevBranchCombinedAucroc) while its
tail component reaches \RevBranchTailOnlyAucroc. No metric computed on the
assembled score can attribute this, and the decomposition that finds it uses
values the scorer already computes.
\item \textbf{A measured batch dependence in a widely used baseline}
(Section~\ref{sec:ecodbatch}). PyOD's ECOD recomputes its empirical CDFs over
the training matrix concatenated with whatever batch is passed to
\texttt{decision\_function}, so a record's score depends on which other records
accompanied it. Holding the evaluated records and the fitted model identical
and varying only the accompanying batch moves ECOD's AUC-PR by
\RevEcodBatchDelta. Published ECOD numbers are therefore not comparable across
studies that score a different batch, in size or composition.
\item \textbf{A provenance discipline that is itself auditable}
(Section~\ref{sec:provenance}), including the correction history and overlap
statement of Section~\ref{sec:disclosure}.
\end{enumerate}

\section{Related Work}
\label{sec:related}

\textbf{Benchmark criticism.} \citet{catillo2023hype} question
whether results on public intrusion datasets represent concrete advances and
warn specifically that partitioning timestamped records in arbitrary order can
alter split representativeness. \citet{arp2022dosdonts} catalogue
sampling bias, mixing incompatible sources and temporal snooping as recurring
security-ML pitfalls. \citet{sommer2010outside} set out
earlier why machine learning is harder to apply to intrusion detection than to
other domains, pointing to the high cost of errors, the semantic gap between
results and operational meaning, and the difficulty of sound evaluation. \citet{ring2019datasets} survey dataset
limitations, and \citet{engelen2021cicids} correct label and
flow-construction defects in CICIDS2017; we build on their corrected release.
\citet{liu2022errorprevalence} quantify how prevalent such labelling
errors are in CIC-IDS-2017 and CSE-CIC-IDS-2018, and \citet{lanvin2022errors} show that correcting the CICIDS2017 errors changes
measured detection performance.
\citet{catillo2021critique} earlier showed that a detector
learned on a public IDS dataset degrades against emulated attacks in a
controlled network, which is the transfer-side counterpart of the question we
ask on the construction side. We do not claim to discover that temporal
assembly is non-neutral. Our addition is quantitative and specific: the same
full multiset under two assemblies, with membership overlap, attack relocation,
prevalence and measured outcome reported together on a real benchmark.

\textbf{The evaluation protocol as the variable under study.} A line of work
treats the protocol itself, rather than the model, as what determines a
reported result. TESSERACT~\citep{pendlebury2019tesseract} shows that spatial
and temporal bias in how train/test sets are constituted inflates malware
classification results, and gives a construction discipline for removing it;
\citet{apruzzese2022crosseval} formalise cross-evaluation and
show how much a reported NIDS result depends on which corpora are combined;
\citet{olszewski2023reproducibility} find that a large fraction
of tier-1 security ML papers cannot be reproduced from what they report.
\citet{apruzzese2023sok} survey how
machine-learning NIDS are assessed and set out pragmatic criteria for
evaluations that practitioners can act on.
Concurrent with this work, \citet{moczkodan2026transformers}
reformulate CICIDS2017 as a temporal task and find that a padding convention,
not the architecture, determines whether Transformers appear to help --- a
protocol variable of a different kind (input formatting) on the same benchmark.
\citet{biswas2026protocol} likewise argues the evaluation protocol is the
hidden variable: within the single corpus CSE-CIC-IDS2018 it compares random,
stratified-temporal and class-blind chronological splits, uses the
stratified-temporal arm to attribute the chronological collapse to attack
families unseen in training rather than to drift within known classes, reports
zero-shot transfer to CIC-DDoS2019 as a separate experiment, and scores each
protocol on its own held-out partition rather than on records held out by
both. This work varies the assembly of one fixed record multiset and
attributes the outcome change to \emph{composition} rather than to
\emph{history order} by restricting both arms to the records they both held
out. The two are complementary: theirs separates unseen-attack coverage from
drift within one corpus and reports transfer between corpora separately, ours
attributes an assembly effect within one corpus under a stated assumption.

\textbf{Streaming anomaly detection.} The baselines evaluated in this rebuild
are Half-Space Trees~\citep{tan2011hst} and LODA~\citep{pevny2016loda}
(streaming), with LOF~\citep{breunig2000lof} and ECOD~\citep{li2023ecod} as batch
diagnostic references. KitNET~\citep{mirsky2018kitsune},
xStream~\citep{manzoor2018xstream}, RRCF~\citep{guha2016rrcf}, streaming
isolation forest~\citep{ding2013iforestasd} and COPOD~\citep{li2020copod} have
implementations in the artifact but no archived evaluation runs in this
rebuild, and carry no numbers here. \citet{cao2024revisiting}
benchmark streaming detectors under controlled anomaly types and drifts and
report strong sensitivity to assessment design; our results give one such
sensitivity a measured mechanism on an IDS benchmark. \citet{wu2023flawed} argue that the standard time-series anomaly
benchmarks are themselves flawed in ways that create an illusion of progress.

\textbf{Change-point detection.} The evaluated detector descends from Bayesian
online change-point detection~\citep{adams2007bocpd} with a truncated run-length
posterior~\citep{fearnhead2007online,knoblauch2018bocpd}.
Section~\ref{sec:detector} shows that the reset posterior of the published
recursion, which the implementation reproduces, is data-independent below the
cap; \citet{vandeburg2020evaluation} document how often change-point
evaluations mismeasure their subject.

\textbf{Conformal and calibrated alerting.} \citet{laxhammar2014conformal} established conformal anomaly detection
for sequential data, with guarantees resting on exchangeability
assumptions~\citep{vovk2005alrw}. Three 2026 lines run concurrent with this work
and are distinct from it. FIRCE~\citep{barrett2026firce} and its successor
FADES~\citep{barrett2026fades} integrate conformal evaluation into streaming IDS
pipelines to drive drift detection and adaptive retraining; both propose
detection machinery and evaluate it, whereas this paper proposes no detector and
instead audits what an evaluation pipeline measures. \citet{gurjar2026tailrisk} train a gradient-boosted classifier on smoothed
alert intensity, momentum and volatility features to predict whether alert
intensity will exceed the training period's 95th percentile within the next
30 minutes, fitted per severity class on 251 million Suricata alerts from one
university network --- a forecasting task on operational alert volume, not a
question about how a benchmark is assembled. None of the three examines the assembly step. Our
results are
relevant but must be stated carefully: reordering one realized dataset does not
establish exchangeability, which is a distributional symmetry. What
interleaving can do is \emph{mask temporal dependence, or make exchangeability
appear more plausible to a diagnostic} applied to the assembled stream.
Conformal methods that relax exchangeability already
exist~\citep{barber2023beyond}.

\textbf{Metric behaviour under class imbalance.} That precision-recall
summaries move with the positive-class base rate is long-established and is not
a finding of this paper. \citet{davis2006prroc} characterise
the ROC/PR correspondence and show that the achievable PR region depends on the
class ratio; \citet{saito2015prplot} demonstrate that PR
summaries are the informative choice precisely because they respond to
imbalance while ROC summaries do not; and \citet{axelsson2000baserate}
established the base-rate consequence for intrusion detection specifically.
\citet{mcdermott2024closerlook} caution that AUC-PR is not
automatically the better summary under class imbalance and that the choice
should follow how the scores will be used. We
use these results, we do not extend them: every AUC-PR this paper compares
\emph{across} prevalence regimes is reported with its chance level $p$ and with
normalized lift $(AP-p)/(1-p)$ for that reason, and comparisons made within one
slice, where the chance level is shared, are reported at that fixed level. What is ours is narrower and is about this benchmark's
construction, not about the metric: the prevalence regime these evaluations
operate in is \emph{doubly constructed} --- first by a retention budget whose
per-day sampling is anti-correlated with attack density, and then by the
assembly step, which redistributes the surviving attacks across the split.
Neither construction step is visible in a reported operating point.

\textbf{Cost-sensitive thresholds.} The threshold actually implemented is the
prior-inclusive cutoff formula of \citet{elkan2001cost}
(Section~\ref{sec:detector}); SRE-style error budgets~\citep{beyer2016sre}
motivated the original framing.

\section{The System Under Study}
\label{sec:detector}

We characterize the detector as implemented, from measurement. All values here
are archived in the artifact.

\subsection{The score the evaluation consumes}
The published description defines the anomaly score as the run-length-reset
posterior $P(r_t{=}0\mid x_{1:t})$. The implementation returns
\[
s_t \;=\; \max\bigl(\text{tail}_t,\; 0.25 \cdot P(r_t \le 5 \mid x_{1:t})\bigr),
\]
where $\text{tail}_t$ is a $\chi^2$ CDF of the squared standardized residual
under a global, slowly adapting diagonal Gaussian. Two facts follow directly.
The evaluated score is not the described score; and it is a function of
$P(r_t\le5)$, so results about $P(r_t{=}0)$ do not transfer to it without
further argument. Because $0.25\cdot P(r\le5)\le0.25$, the auxiliary branch can
set the score only where the tail term is below $0.25$. Measured over the
\SThreeBindingRows{} post-warm-up records at the \emph{head} of the
timestamp-ordered CICIDS2017 stream --- a prefix, not the whole stream, and one
that carries no attacks at all --- it does so on
\SThreeChangePointBindingPct\%\ of records, at a mean contributed value of
\SThreeChangePointBindingMeanValue. The same share measured on the
attack-bearing held-out slice is \RevBranchAuxBindsHeldoutPct\%\
(Section~\ref{sec:branch}); the two are different populations and are not in
tension. A mean magnitude does
not establish which branch \emph{discriminates}, since ranking metrics are
invariant to the scale of a branch that sets the ranking; we therefore make no
claim about that here and measure it directly in
Section~\ref{sec:branch}.

\subsection{The run-length posterior below the cap, and at it}
In the recursion of \citet{adams2007bocpd} as published
(their Eq.~3 and Algorithm~1), the changepoint and growth branches at step $t$
carry the same run-conditional predictive term, which cancels in the posterior normalization: under a
constant hazard the reset posterior $P(r_t{=}0\mid x_{1:t})$ equals the hazard
rate $h$ \emph{exactly, for any data}, because the reset hypothesis at $t$ is
not informed by $x_t$ and a change shows in the posterior one step later, on
$r_{t+1}{=}1$. The implementation reproduces that recursion below the
run-length cap, so this is a property of the algorithm, not a coding error.
The audit finding is narrower: the earlier description defined the score as
that reset posterior, which is uninformative below the cap by construction,
whereas the implemented score is the $\max$ composition over $P(r_t\le5)$
stated above. Under a $6\sigma$ mean shift with
hazard \SThreeProbeHazard: mean $P(r{=}0)$ before the shift
\SThreePrZeroPreMean; peak in the 50-record window after it
\SThreePrZeroResponsePeak; maximum deviation from the hazard between
initialization and truncation \SThreePrZeroSteadyMaxDeviation.

That exact result covers only part of the evaluated regime, and we state the
limit plainly. At $t\ge\SThreeRunLengthCap$ the run array is truncated before
normalization, the cancellation no longer holds, and the posterior wanders
(mean \SThreePrZeroTruncatedMean, maximum \SThreePrZeroTruncatedMax). Every
stream evaluated in this paper is far longer than \SThreeRunLengthCap{}
records, so the published runs spent nearly all of their length in the
truncated regime. We have therefore proved data-independence where it is
provable and measured behaviour where it is not; we have not established that
the auxiliary branch is uninformative over the regime that produced almost all
reported scores, and we do not claim it.

\subsection{The threshold}
The implemented decision cutoff is the prior-inclusive formula of
\citet{elkan2001cost},
$\tau = C_{FP}(1-\rho)/[C_{FP}(1-\rho) + C_{FN}\rho]$, giving
\SThreeThresholdLitnet{} at incident prior $\rho{=}0.05$ and
\SThreeThresholdCicids{} at $\rho{=}0.22$ --- not the cost-only
$C_{FP}/(C_{FP}{+}C_{FN}) = \SThreeThresholdCostOnly$ described in earlier
versions. That formula is a Bayes rule for a calibrated class posterior. The
quantity it is applied to here is the $\max$ composition defined at the head of
this section, which is not one, so the cutoff carries no Bayes-optimality
warrant on this score. We record that as an observation about what the
implementation does and not as a defect claim, and it changes no reported
number: AUC-PR and AUC-ROC are threshold-free and unaffected; we report no
threshold-dependent column, which is also why none of the non-comparable
precision/recall/F1 columns appears in this paper.

\subsection{Detection delay, not latency}
The quantity previously reported as ``latency in milliseconds'' is a count of
records between attack onset and first alert; the evaluation contains no
wall-clock instrumentation (\SThreeLatencyUsesAClock{} clock calls in the
module). We report it as detection delay in records and report no per-flow
compute cost, which this pipeline does not measure.

\paragraph{An alternative reset formulation, in brief.}
Appendix~\ref{app:ablation} reports one untuned instance of a formulation that
places a prior-predictive term on the reset branch, so that the reset
hypothesis is informed by $x_t$. Measured under a 30-minute compute cap on one
LITNET stream, one \SSixPrefixRows-record prefix and one seed, it responds on
the synthetic probe ($P(r{=}0)$ peaks at \SSixProbeCorrectedPeak{} against
\SSixProbeOriginalPeak{} for the evaluated detector) but ranks near chance on
real data ($AP=\SSixCorrectedAucpr$ against \SSixOriginalAucpr, AUC-ROC
\SSixCorrectedAucroc): its short-run mass saturates and the score sits at the
$0.25$ cap on \SSixDiagCorrectedFracAtCap{} of the diagnostic records. The two
variants are degenerate in different quantities, and we claim neither as
change-point detection; the appendix states each in the quantity measured, and
what the one-stream, one-seed run does not establish.

\section{Datasets, Assembly, and Protocol}
\label{sec:streams}

\subsection{CICIDS2017: a timestamp-sorted budgeted subsample}
We use the Engelen-corrected release~\citep{engelen2021cicids}. The evaluation
corpus is a \textbf{timestamp-sorted budgeted subsample}: \CicidsWeekUsedRows{}
of \CicidsWeekEligibleRows{} eligible records, a retention of
\CicidsWeekRetentionPct\%, selected by fixed per-day budgets with proportional
stride subsampling. Because the budgets are fixed rather than a common factor,
larger days are cut harder and the attack-heavy days are the largest: per-day
retention runs from \CicidsWeekMinDayRetentionPct\%\ (Friday, the densest day)
to \CicidsWeekMaxDayRetentionPct\%\ (Tuesday). Measured consequence: the
eligible week's prevalence is \CicidsWeekTruePrevalencePct\%\ while the corpus
is \CicidsWeekUsedPrevalencePct\%, a bias of \CicidsWeekPrevalenceBiasPp{}
points from the budget step alone. Sorting a stride-subsampled corpus preserves
relative order but is not the full event stream; missing flows can break attack
runs and alter online sufficient statistics. We measured the prevalence bias
and did not measure the effect on attack-family mix, feature distributions or
run lengths, so we use ``timestamp-sorted budgeted subsample'' throughout and
avoid claims that require the complete natural chronology.
Table~\ref{tab:streams} summarizes all four streams.

\begin{table}
\caption{The evaluation streams (Stage 1 manifests). Run lengths are contiguous
attack runs in the labelled stream; span is the wall-clock extent of the
capture.}
\label{tab:streams}
\begin{tabular}{lrrrrr}
\toprule
stream & records & prevalence & held-out prev. & run med/p90/max & span (min) \\
\midrule
CICIDS2017 (week) & \StreamCicidsTwoZeroOneSevenRows & \StreamCicidsTwoZeroOneSevenPrevalencePct\% & \StreamCicidsTwoZeroOneSevenTestPrevalencePct\% & \StreamCicidsTwoZeroOneSevenRunMedian/\StreamCicidsTwoZeroOneSevenRunPninety/\StreamCicidsTwoZeroOneSevenRunMax & \StreamCicidsTwoZeroOneSevenSpanMinutes \\
LITNET udp\_flood & \StreamLitnetUdpFloodRows & \StreamLitnetUdpFloodPrevalencePct\% & \StreamLitnetUdpFloodTestPrevalencePct\% & \StreamLitnetUdpFloodRunMedian/\StreamLitnetUdpFloodRunPninety/\StreamLitnetUdpFloodRunMax & \StreamLitnetUdpFloodSpanMinutes \\
LITNET blaster\_worm & \StreamLitnetBlasterWormRows & \StreamLitnetBlasterWormPrevalencePct\% & \StreamLitnetBlasterWormTestPrevalencePct\% & \StreamLitnetBlasterWormRunMedian/\StreamLitnetBlasterWormRunPninety/\StreamLitnetBlasterWormRunMax & \StreamLitnetBlasterWormSpanMinutes \\
LITNET spam & \StreamLitnetSpamRows & \StreamLitnetSpamPrevalencePct\% & \StreamLitnetSpamTestPrevalencePct\% & \StreamLitnetSpamRunMedian/\StreamLitnetSpamRunPninety/\StreamLitnetSpamRunMax & \StreamLitnetSpamSpanMinutes \\
\bottomrule
\end{tabular}
\end{table}

\subsection{LITNET-2020: three disjoint attack-type captures}
The three captures this paper evaluates have no common chronology: they do not
overlap in time, they are separated by gaps of weeks and months, and their
spans are \StreamLitnetUdpFloodSpanMinutes, \StreamLitnetBlasterWormSpanMinutes{}
and \StreamLitnetSpamSpanMinutes{} minutes (Table~\ref{tab:streams}), so a
global timestamp order over them would concatenate three intervals across
those gaps rather than describe one observed stream. We make no chronology
claim about LITNET-2020 as a whole, which was collected over about ten months
and annotates twelve attack types~\citep{damasevicius2020litnet}. We evaluate three per-attack-type
streams of \StreamLitnetUdpFloodRows{} records each, never a manufactured
composite; the pooled composite appears only as the explicitly labelled
assembled arm of Section~\ref{sec:contrast-litnet}. Multi-window burn-rate
alerting over 60/360/4320-minute windows is undefined on the two captures that
span minutes; we report that as a finding about the benchmark, and no burn-rate
result appears in this paper.

\subsection{Protocol}
\label{sec:protocol}
Tables~\ref{tab:protocol} and~\ref{tab:protocol-baselines} state the protocol
in full so the paper is self-contained; the artifact reproduces it exactly.

\begin{table}
\caption{Experimental protocol. Every entry is taken from the code the archived
manifests pin, not from intent.}
\label{tab:protocol}
\begin{tabular}{@{}lp{0.72\linewidth}@{}}
\toprule
element & setting \\
\midrule
features & all numeric columns after dropping the label; non-numeric columns
dropped, so no categorical encoding is used; $d=\ContrastCicidsDim$ (CICIDS2017),
\ContrastLitnetUdpFloodDim{} (udp\_flood), \ContrastLitnetBlasterWormDim{}
(blaster\_worm), \ContrastLitnetSpamDim{} (spam) \\
scaling & none applied by the harness \\
missing / infinite & $\pm\infty$ mapped to NaN, then all NaN filled with $0.0$;
this touches \RefImputeCicidsAffectedRows{} of the \RefImputeCicidsRows{}
CICIDS2017 rows (\RefImputeCicidsInfCells{} infinite cells in
\RefImputeCicidsAffectedFeatures{} features, no NaN) and
\RefImputeLitnetUdpFloodAffectedRows{}, \RefImputeLitnetBlasterWormAffectedRows{}
and \RefImputeLitnetSpamAffectedRows{} rows of the three LITNET streams \\
label rule & $y{=}1$ unless the label string is \texttt{benign}, \texttt{normal},
\texttt{0} or \texttt{false} \\
eligibility (CICIDS) & rows whose label contains `` - Attempted'' dropped \\
per-day budgets & 300k / 300k / 350k / 300k / 350k (Mon--Fri) \\
subsampling & proportional stride within each class, preserving per-day
prevalence \\
split & positional 70/15/15 over the assembled stream order \\
detector: hazard & \SThreeProbeHazard \\
detector: run-length cap & \SThreeRunLengthCap \\
detector: warm-up & 30 records \\
detector: short-run mass & $P(r\le5)$, weight $0.25$ \\
detector: predictive model & one diagonal Gaussian over all $d$ features,
initialised on the first record, with mean and $M_2$ advanced by Welford's
recurrence after every record and variances floored at $10^{-4}$; the tail
score is the $\chi^2_d$ CDF of the squared standardized residual under the
pre-update global statistics, zero during warm-up; run-conditional means and
$M_2$ are kept per run length by the same recurrence \\
metric & \texttt{sklearn.metrics.average\_precision\_score} (AUC-PR) and
\texttt{roc\_auc\_score}, scikit-learn 1.8.0, positive class $1$, no sample
weighting; average precision is the step-wise sum
$\sum_n (R_n-R_{n-1})P_n$, not trapezoidal interpolation. Flow-wise average
precision is the convention of the compared work; event-based precision and
recall for time series~\citep{tatbul2018precision} are future work \\
chance level & the expected average precision of an uninformative ranking is
taken as the positive-class prevalence $p$; the exact finite-sample expectation
of step-wise $AP$ under a uniformly random permutation differs from $p$ in the
fifth decimal (\ChanceExactNaturalSliceAucpr{} against
\SFourCicidsNaturalChanceFloor{} on the timestamp-order held-out slice; the
full-list case of Theorem~1 of \citealt{manzhos2026ap}) and
changes no conclusion here; it is an expectation, not a lower bound, and a
ranking can fall below it \\
reported precision & every \emph{dimensionless} metric on $[0,1]$ ---
AUC-PR, AUC-ROC, chance level, fractional prevalence, normalized and additive
lift, and the margins, ranges and differences taken between them, including
per-seed draws --- is reported to six decimal places, and any of those derived
from other reported quantities is computed from their \emph{reported} values.
Every such derived number therefore equals the arithmetic on the numbers
printed beside it, exactly, and a reader can check any of them; the cost is
that it may differ from its full-precision counterpart by a few units in the
last reported place, below $10^{-5}$ throughout and orders of magnitude below
any effect claimed here. Deliberately outside that rule, and reported at the
precision each was emitted with: percentage- and percentage-point-scaled
values (the same quantities at $100\times$, where six decimals would fabricate
digits), integer counts, the protocol constants in this table and in
Table~\ref{tab:protocol-baselines}, the posterior
probabilities of Section~\ref{sec:detector} and the diagnostic fractions of
Table~\ref{tab:ablation}, which are not metrics, and values quoted from cited
work \\
\bottomrule
\end{tabular}
\end{table}

\begin{table}
\caption{Experimental protocol, continued: update timing, label access,
baseline hyperparameters, seeds and the sweep's resampling procedure. Every
entry is read from the code the archived manifests pin.}
\label{tab:protocol-baselines}
\begin{tabular}{@{}lp{0.72\linewidth}@{}}
\toprule
element & setting \\
\midrule
update timing & prequential, in the sense of \citet{gama2013prequential}: score computed from the pre-update predictive;
sufficient statistics updated after scoring, on every record including held-out
ones \\
label access & the detector and HST see no labels at any point; baselines
receive a validation-derived threshold; \textbf{ECOD and LOF are fitted on benign-only
training rows and therefore have label access} \\
decision threshold & the evaluated detector alerts at Elkan's prior-inclusive
cutoff $\tau = C_{FP}(1-\rho)/[C_{FP}(1-\rho)+C_{FN}\rho]$ with $C_{FP}=1$,
$C_{FN}=10$ and $\rho=0.05$ in the contrast runs (\SThreeThresholdLitnet);
each baseline alerts at the threshold that maximises F1 on the validation
slice's precision--recall curve, falling back to the same $\tau$ when
validation holds one class; AUC-PR and AUC-ROC are threshold-free and no
threshold-dependent column is reported \\
HST & river \texttt{HalfSpaceTrees} behind a river \texttt{MinMaxScaler}:
25 trees, height 15, window 250, the cell's seed \\
LODA & the artifact's own implementation of LODA: 100 unit-norm random
projections, 32 histogram bins, window 512, running standardization, the
cell's seed \\
LOF & \texttt{sklearn.neighbors.LocalOutlierFactor}, 35 neighbours,
\texttt{novelty=True}, \texttt{contamination='auto'}, fitted on benign-only
training rows; score $=-$\texttt{score\_samples}, min--max scaled \\
ECOD & PyOD 2.0.5 \texttt{ECOD()} at its defaults, fitted on benign-only
training rows and scored in one \texttt{decision\_function} call on the batch stated in
Section~\ref{sec:ecodbatch}; scores min--max scaled \\
seeds & contrast and LITNET runs: seed 11, with HST's additional draws at seeds
23 and 47; sweep: resample seeds 11, 23 and 47, each streaming method seeded
with the resample seed \\
sweep resampling & stratified within each chronological split segment: below
the natural rate every benign row is kept and attacks are thinned to the
segment quota; above it both classes are thinned to the quotas of the largest
stream whose re-split meets the target in every segment; selection is random
without replacement, order preserved, no duplicates; a draw is accepted only if
every segment's achieved prevalence is within 1\,pp of the target, otherwise
redrawn with seed $+\,100000\cdot$attempt, at most 20 redraws \\
\bottomrule
\end{tabular}
\end{table}

\noindent\begin{minipage}{\linewidth}
Stream assembly, stated as an algorithm:
{\footnotesize
\begin{verbatim}
timestamp-order arm:
  rows <- corpus sorted by Timestamp (stable mergesort)
day-round-robin arm:
  groups <- rows grouped by capture date, each in timestamp order
  for k = 0, 1, 2, ...:
      for d in sorted(dates):
          if k < len(groups[d]): emit groups[d][k]
pooled-composite arm (LITNET):
  groups <- per-attack-type streams, each in timestamp order
  emit round robin over sorted(attack_type) as above
split (both arms): i_tr = floor(0.70 n); i_va = i_tr + floor(0.15 n)
\end{verbatim}
}
\end{minipage}

\section{Assembly as a Treatment}
\label{sec:contrast}

\subsection{Design and what it identifies}
CICIDS2017 admits a clean manipulation of the assembly step: the same record
multiset can be presented in timestamp order or in day-of-week round robin. The
reordering is verified mechanically to be an exact permutation --- both arms
hold \CicidsArmRows{} records and \CicidsArmAttacks{} attacks
(\CicidsArmPrevalencePct\%). The assembled arm reproduces the archived earlier
evaluation exactly (\SFourCicidsInterleavedTestAttacks{} held-out attacks in
\CicidsHeldoutSize{} records), so the contrast is against the assembly actually
used, not a strawman.

\begin{figure}
\centering
\includegraphics[width=\textwidth]{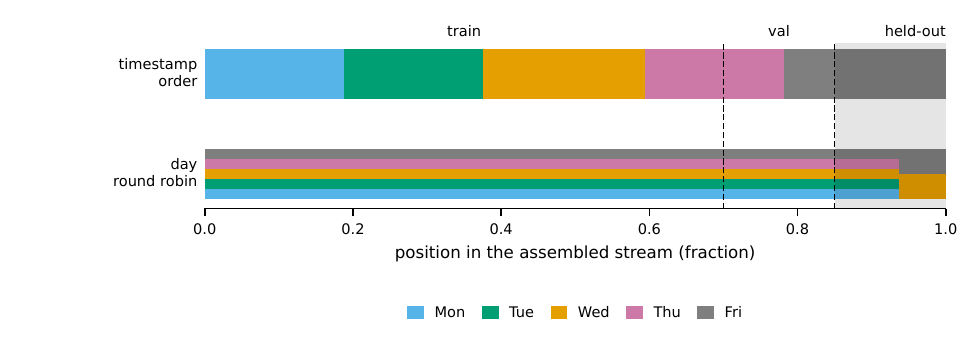}
\caption{The two assemblies of the same CICIDS2017 records under one fixed
positional split. In timestamp order the capture days follow one another, so
the held-out slice is the tail of the last day. In day round robin every day
still supplying records contributes to every position, so the held-out slice
draws from the tail of each day. Block widths follow the per-day record
budgets of Table~\ref{tab:protocol}. The figure is a schematic of the
protocol and carries no measured value.}
\label{fig:assembly}
\end{figure}

What the design does \emph{not} do is hold the evaluated sample fixed. Under a
positional split, reordering changes split membership, so the two arms differ
in held-out records, held-out prevalence, training composition and within-split
order at once. The estimand is therefore the effect of the complete assembly
pipeline under a fixed split rule, and nothing finer. We report the joint change
and decline to decompose it. Figure~\ref{fig:assembly} shows why the held-out
membership differs.

\subsection{What assembly does to the evaluated sample}
Held-out prevalence differs by \SFourCicidsPrevShiftAbsPp{} points. The two
held-out slices share only \CicidsHeldoutOverlap{} of \CicidsHeldoutSize{}
records (\CicidsHeldoutOverlapPct\%); the assembled arm's held-out attacks are
a strict subset of the timestamp-ordered arm's, with
\CicidsAttacksMovedOutOfHeldout{} attacks relocated out of the held-out slice:
\RefRelocatedToTraining{} of them into training and \RefRelocatedToValidation{}
into validation, counted from the archived splits. The
mechanism of the prevalence change is dilution: every held-out attack in the
assembled arm is a Friday record
(\CicidsSyntheticHeldoutAttackDays{} of five days contribute any), while
Monday--Thursday contribute \CicidsHeldoutAttackFreeRows{} attack-free held-out
rows. The assembled arm's Friday sub-slice is denser
(\CicidsFridayHeldoutDensityPct\%) than the timestamp-ordered arm's entire
slice: prevalence falls because attack-free rows from the four other days enter
the held-out slice and displace Friday attack rows, which is the same movement
the relocation counts above report. This dilution accounts for the prevalence change; it
is not by itself an account of the outcome change in
Section~\ref{sec:outcome}.

\begin{table}
\caption{The assembly contrast. Held-out prevalence, attacks, and the evaluated
detector's AUC-PR per arm. Every cell is a macro resolving to an archived
manifest.}
\label{tab:contrast}
{\small
\begin{tabular}{llrrr}
\toprule
Benchmark & Construction & Held-out prev. & Attacks & AUC-PR (evaluated detector) \\
\midrule
CICIDS2017 & natural (timestamp) & \SFourCicidsNaturalTestPrev\% & \SFourCicidsNaturalTestAttacks & \SFourCicidsNaturalProposedDetectorAucpr \\
CICIDS2017 & synthetic (day RR) & \SFourCicidsInterleavedTestPrev\% & \SFourCicidsInterleavedTestAttacks & \SFourCicidsSyntheticProposedDetectorAucpr \\
LITNET udp flood & natural (per type) & \SFourLitnetUdpFloodTestPrev\% & \SFourLitnetUdpFloodTestAttacks & \SFourLitnetUdpFloodProposedDetectorAucpr \\
LITNET blaster worm & natural (per type) & \SFourLitnetBlasterWormTestPrev\% & \SFourLitnetBlasterWormTestAttacks & \SFourLitnetBlasterWormProposedDetectorAucpr \\
LITNET spam & natural (per type) & \SFourLitnetSpamTestPrev\% & \SFourLitnetSpamTestAttacks & \SFourLitnetSpamProposedDetectorAucpr \\
LITNET pooled & synthetic (3-type RR) & \SFourLitnetPooledTestPrev\% & \SFourLitnetPooledTestAttacks & \SFourLitnetPooledProposedDetectorAucpr \\
\bottomrule
\end{tabular}
}
\end{table}

\subsection{What assembly does to the measured outcome}
\label{sec:outcome}
AUC-PR's chance level is the positive-class prevalence, which the treatment
moves (\SFourCicidsNaturalChanceFloor{} against
\SFourCicidsSyntheticChanceFloor), and additive lift $AP-p$ has a
prevalence-dependent ceiling of $1-p$. We therefore report raw $AP$, the chance
level $p$, and the normalized lift $(AP-p)/(1-p)$ together, and report no raw
cross-arm difference.

\begin{itemize}
\item \textbf{Timestamp order} ($p=\SFourCicidsNaturalChanceFloor$): ECOD
$AP=\SFourCicidsNaturalEcodAucpr$ (normalized lift
\SFourCicidsNaturalEcodNormLift), detector
$AP=\SFourCicidsNaturalProposedDetectorAucpr$ (normalized lift
\SFourCicidsNaturalProposedDetectorNormLift). HST scores
\SFourCicidsNaturalHstAucpr{} here, \SFourCicidsNaturalHstLiftAbs{} below its
chance level, from a single seed, so no placement is asserted for it.
\item \textbf{Day round robin} ($p=\SFourCicidsSyntheticChanceFloor$): detector
$AP=\SFourCicidsSyntheticProposedDetectorAucpr$ (normalized lift
\SFourCicidsSyntheticProposedDetectorNormLift), ECOD
$AP=\SFourCicidsSyntheticEcodAucpr$ (normalized lift
\SFourCicidsSyntheticEcodNormLift). HST's three-seed values are
\SFourSeedCicidsTwoZeroOneSevenInterleavedSyntheticHstDrawOne,
\SFourSeedCicidsTwoZeroOneSevenInterleavedSyntheticHstDrawTwo{} and
\SFourSeedCicidsTwoZeroOneSevenInterleavedSyntheticHstDrawThree{} (mean
\SFourSeedCicidsTwoZeroOneSevenInterleavedSyntheticHstMean, sd
\SFourSeedCicidsTwoZeroOneSevenInterleavedSyntheticHstSd); its ordering against
ECOD flips across those seeds, so it carries no placement.
\end{itemize}

\begin{figure}
\centering
\includegraphics[width=\textwidth]{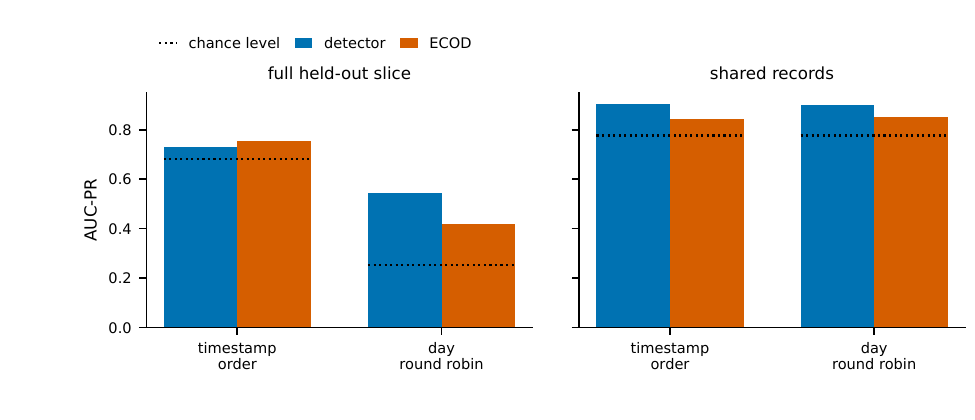}
\caption{Detector and ECOD AUC-PR in the two arms. Left: the full held-out
slice of each arm, where the measured ordering reverses. Right: the
\RevSharedRecords{} records both arms held out, where the detector leads in
both arms; the ECOD bars there are the slice-batch values of
Table~\ref{tab:shared}. Dotted marks are each group's chance level, the held-out
prevalence: \SFourCicidsNaturalChanceFloor{} and
\SFourCicidsSyntheticChanceFloor{} for the full slices and
\RevSharedPrevalence{} for the shared records. Values are those of
Section~\ref{sec:outcome} and Table~\ref{tab:shared}.}
\label{fig:decomposition}
\end{figure}

The measured ordering of the two deterministic scorers reverses between arms,
in both raw and normalized terms. \textbf{This is a pipeline-level evaluation
outcome under the stated protocol, not a statement about method quality.} Three
qualifications are load-bearing and are stated here rather than deferred.
(i)~ECOD is fitted on benign-only training rows and therefore has label access
the streaming methods do not; we treat it throughout as a \emph{label-privileged
diagnostic reference}, and no superiority is claimed for or against it.
(ii)~The two arms score different held-out samples, so the reversal is not
measured on a common sample. Section~\ref{sec:shared} measures what happens
when that difference is removed, and the reversal does not survive it
(Figure~\ref{fig:decomposition}).
(iii)~Determinism removes seed variance but not sampling uncertainty from the
selected window, correlated attack runs, or the subsample. The margins are
\SFourCicidsNaturalEcodMinusDetectorAucpr{} $AP$ for ECOD in timestamp order
and \SFourCicidsSyntheticDetectorMinusEcodAucpr{} $AP$ for the detector in day
round robin, each computed from the reported cells; their event-block bootstrap
intervals over the held-out slice (Section~\ref{sec:threats} states the
procedure; the intervals are computed from the archived per-record scores) are
$[\RefBootFullNaturalMarginLo,\,\RefBootFullNaturalMarginHi]$ and
$[\RefBootFullSyntheticMarginLo,\,\RefBootFullSyntheticMarginHi]$, neither of
which crosses zero. Those intervals cover sampling uncertainty over this slice
only, so we claim no stable ordering, only a measured one under this protocol.

\subsection{LITNET-2020: pooling as an audit check}
\label{sec:contrast-litnet}
Round robin \emph{within} a per-attack-type stream is the identity, so the
LITNET contrast is compositional: three per-type streams against the pooled
composite used by earlier versions. The pooled held-out slice is by
construction exactly the union of the three per-type held-out slices --- equal
budgets, a perfect three-cycle, and a divisible split boundary --- so its
prevalence \SFourLitnetPooledTestPrev\%\ is the equal-weight mean of
\SFourLitnetUdpFloodTestPrev\%, \SFourLitnetBlasterWormTestPrev\%\ and
\SFourLitnetSpamTestPrev\%. This is arithmetic, and we present it as an audit
check with pedagogical value rather than as a finding: a pooled operating point
can correspond to no capture in the pool, and the equal weighting is an
artifact of equal budgets. Per-stream values are in Table~\ref{tab:litnet}.

\begin{table}
\caption{LITNET-2020: AUC-PR per stream against the pooled assembled composite.
The detector and ECOD (label-privileged diagnostic reference) are deterministic;
HST is a single draw per cell, so no ordering involving HST is asserted. Its
three composite draws across seeds are
\SFourSeedLitnetPooledCompositeSyntheticHstDrawOne,
\SFourSeedLitnetPooledCompositeSyntheticHstDrawTwo,
\SFourSeedLitnetPooledCompositeSyntheticHstDrawThree.}
\label{tab:litnet}
\begin{tabular}{lrrrr}
\toprule
stream & held-out prev. & detector & ECOD & HST (1 seed) \\
\midrule
udp\_flood (per type) & \SFourLitnetUdpFloodTestPrev\% & \SFourLitnetUdpFloodProposedDetectorAucpr & \SFourLitnetUdpFloodEcodAucpr & \SFourLitnetUdpFloodHstAucpr \\
blaster\_worm (per type) & \SFourLitnetBlasterWormTestPrev\% & \SFourLitnetBlasterWormProposedDetectorAucpr & \SFourLitnetBlasterWormEcodAucpr & \SFourLitnetBlasterWormHstAucpr \\
spam (per type) & \SFourLitnetSpamTestPrev\% & \SFourLitnetSpamProposedDetectorAucpr & \SFourLitnetSpamEcodAucpr & \SFourLitnetSpamHstAucpr \\
pooled (assembled) & \SFourLitnetPooledTestPrev\% & \SFourLitnetPooledProposedDetectorAucpr & \SFourLitnetPooledEcodAucpr & \SFourLitnetPooledHstAucpr \\
\bottomrule
\end{tabular}
\end{table}

\section{Isolating What the Treatment Changes}
\label{sec:shared}

The contrast of Section~\ref{sec:contrast} moves membership, prevalence and
order together. Three analyses separate what can be separated from the archived
per-record scores of both arms. All three use the metric implementation of
Table~\ref{tab:protocol}; the dumps reproduce the archived arms to
\RevReproNaturalDelta{} AUC-PR on the timestamp-ordered arm and exactly on the
assembled arm.

\subsection{The records both arms held out}
The two held-out slices share \RevSharedRecords{} records, carrying
\RevSharedAttacks{} attacks at prevalence \RevSharedPrevalence{} --- they are
the Friday records both assemblies happen to place in the test set, which is
why their prevalence is far above either arm's. Restricting both arms to
exactly those records, and recomputing:

\begin{table}
\caption{Both arms restricted to the \RevSharedRecords{} records they both
held out. Membership and prevalence are held fixed by construction;
everything the assembly changes upstream of scoring still varies: the
detector's prefix, ECOD's benign-only training rows and, in the slice
columns, its batch. None of it reverses the ordering: the detector moves by
\RevSharedDetectorSpread{} $AP$ between the two prefixes and ECOD by
\RefEcodSharedSevenEightCrossArmSpread{} $AP$ between the two fitted models on
the identical shared batch. ECOD is reported under
two scoring batches: ``slice'' is the arm's full \CicidsHeldoutSize-record
held-out slice, scored together and restricted afterwards to the shared
records; ``shared'' is the \RevSharedRecords{} shared records scored as their
own batch under the same fitted model, which is the batch-controlled
comparison (Section~\ref{sec:ecodbatch}).}
\label{tab:shared}
{\small
\begin{tabular}{lrrrrrr}
\toprule
 & \multicolumn{3}{c}{$AP$} & \multicolumn{3}{c}{AUC-ROC} \\
\cmidrule(lr){2-4}\cmidrule(lr){5-7}
arm & detector & ECOD (slice) & ECOD (shared) & detector & ECOD (slice) & ECOD (shared) \\
\midrule
timestamp order & \RevSharedDetectorNaturalAucpr & \RevSharedEcodNaturalAucpr & \RefEcodSharedSevenEightNaturalAucpr & \RevSharedDetectorNaturalAucroc & \RevSharedEcodNaturalAucroc & \RefEcodSharedSevenEightNaturalAucroc \\
day round robin & \RevSharedDetectorSyntheticAucpr & \RevSharedEcodSyntheticAucpr & \RefEcodSharedSevenEightSyntheticAucpr & \RevSharedDetectorSyntheticAucroc & \RevSharedEcodSyntheticAucroc & \RefEcodSharedSevenEightSyntheticAucroc \\
\bottomrule
\end{tabular}}
\end{table}

\textbf{The ordering does not reverse on the shared records, under either
batch.} The detector leads ECOD in both arms: by \RevSharedMarginNatural{} and
\RevSharedMarginSynthetic{} $AP$ under the slice batch and by
\RefMarginSharedSevenEightNatural{} and \RefMarginSharedSevenEightSynthetic{} $AP$ under the
shared batch, each computed from the reported cells of
Table~\ref{tab:shared}, with event-block
bootstrap intervals (Section~\ref{sec:threats})
$[\RefBootSharedTwoFourZeroNaturalMarginLo,\,\RefBootSharedTwoFourZeroNaturalMarginHi]$,
$[\RefBootSharedTwoFourZeroSyntheticMarginLo,\,\RefBootSharedTwoFourZeroSyntheticMarginHi]$,
$[\RefBootSharedSevenEightNaturalMarginLo,\,\RefBootSharedSevenEightNaturalMarginHi]$ and
$[\RefBootSharedSevenEightSyntheticMarginLo,\,\RefBootSharedSevenEightSyntheticMarginHi]$, none
of which crosses zero. The detector's own score changes by only
\RevSharedDetectorSpread{} $AP$ between the two histories, and ECOD's
batch-controlled score by \RefEcodSharedSevenEightCrossArmSpread{} $AP$ between the two
fitted models. This is the paper's most consequential result and it constrains
the paper's central claim: the reversal reported in Section~\ref{sec:outcome}
is attributable to \emph{which records the assembly places in the test set},
not to the order in which the detector processed its history. That attribution
rests on an assumption the archived design does not test: that history
contributes no more on the records the arms do not share than on those they
do. Stated exactly, the restriction shows that the ordering does not reverse on
a sample both arms held out and that the detector's score on that sample moves
by \RevSharedDetectorSpread{} $AP$ between the two histories; it does not bound
history's effect on comparisons between shared and unshared records, which the
full-slice $AP$ of Section~\ref{sec:outcome} also contains. We report the
reversal as an evaluation-level consequence of assembly and
explicitly not as an effect of ordering on detector behaviour.

Two cautions on reading Table~\ref{tab:shared}. The shared records are not a
random subsample --- they are the intersection two particular assemblies
produce, at a prevalence unlike either arm's --- so its absolute values
characterise that intersection, not the benchmark. And the two rows still
differ in everything upstream of scoring, the detector's prefix and ECOD's
benign-only training rows; what is held fixed is the evaluated sample, not the
model state.

\subsection{Where the split falls}
\begin{figure}
\centering
\includegraphics[width=\textwidth]{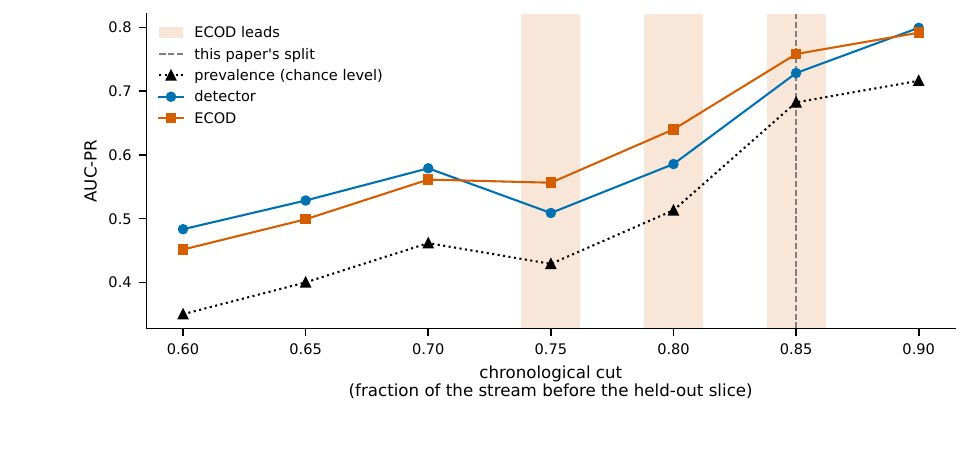}
\caption{AUC-PR of the two deterministic scorers at each of the
\RevSplitCuts{} chronological cuts of the timestamp-ordered CICIDS2017 stream.
The held-out prevalence at each cut is drawn on the same axis. It is the
AUC-PR chance level. Shaded cuts are the \RevSplitEcodWins{} at which ECOD
leads. The dashed line marks the cut this paper's fixed split uses. Values are
recomputed from the archived per-record scores, as those of
Table~\ref{tab:branch} are, with ECOD refit at each cut, and so differ slightly
from Table~\ref{tab:contrast} at the cut this paper uses.}
\label{fig:split}
\end{figure}

A detector score depends only on the records preceding it, so one pass supports
every chronological cut point; ECOD is refit at each cut because its training
set moves with the boundary. Over \RevSplitCuts{} cuts from 60\%\ to 90\%\ of
the stream, ECOD exceeds the detector at \RevSplitEcodWins{} of them, including
the 85\%\ cut this paper's fixed split uses, and the detector leads at the
others. The measured ordering is therefore a property of the split point as well
as of the assembly; the archived split is not adversarially chosen, but it is
not neutral either, and no claim here is a stable property of CICIDS2017.
Per-cut values are in Table~\ref{tab:pairedcuts} and Figure~\ref{fig:split}.
Repeating the assembly contrast at the same \RefPairedCuts{} cuts, with both
arms scored at each cut and ECOD refit at each, the two arms' orderings differ
at \RefPairedCutsOrderingsDiffer{} cuts --- 75\%, 80\%\ and the 85\%\ cut this
paper uses, where ECOD leads in timestamp order and the detector leads in day
round robin --- and agree at the other \RefPairedCutsOrderingsAgree, where the
detector leads in both arms; the detector leads in the day-round-robin arm at
every cut. The reversal is therefore a property of this region of cuts, not of
every cut.

\begin{table}
\caption{The paired cut-by-assembly sweep: chance level (held-out prevalence)
and AUC-PR of both scorers in both arms at the \RefPairedCuts{} chronological
cuts, recomputed from the archived per-record scores with ECOD refit at each
cut; ``agree'' marks the cuts at which the two arms order the scorers the same
way. Figure~\ref{fig:split} draws the timestamp-order columns.}
\label{tab:pairedcuts}
\begin{tabular}{lrrrrrrc}
\toprule
 & \multicolumn{3}{c}{timestamp order} & \multicolumn{3}{c}{day round robin} & \\
\cmidrule(lr){2-4}\cmidrule(lr){5-7}
cut & chance & detector & ECOD & chance & detector & ECOD & agree \\
\midrule
60\% & \RefPairedCutSixZeroNaturalPrevalence & \RefPairedCutSixZeroNaturalDetectorAucpr & \RefPairedCutSixZeroNaturalEcodAucpr & \RefPairedCutSixZeroSyntheticPrevalence & \RefPairedCutSixZeroSyntheticDetectorAucpr & \RefPairedCutSixZeroSyntheticEcodAucpr & \ifnum\RefPairedCutSixZeroOrderingsAgree=1 yes\else no\fi \\
65\% & \RefPairedCutSixFiveNaturalPrevalence & \RefPairedCutSixFiveNaturalDetectorAucpr & \RefPairedCutSixFiveNaturalEcodAucpr & \RefPairedCutSixFiveSyntheticPrevalence & \RefPairedCutSixFiveSyntheticDetectorAucpr & \RefPairedCutSixFiveSyntheticEcodAucpr & \ifnum\RefPairedCutSixFiveOrderingsAgree=1 yes\else no\fi \\
70\% & \RefPairedCutSevenZeroNaturalPrevalence & \RefPairedCutSevenZeroNaturalDetectorAucpr & \RefPairedCutSevenZeroNaturalEcodAucpr & \RefPairedCutSevenZeroSyntheticPrevalence & \RefPairedCutSevenZeroSyntheticDetectorAucpr & \RefPairedCutSevenZeroSyntheticEcodAucpr & \ifnum\RefPairedCutSevenZeroOrderingsAgree=1 yes\else no\fi \\
75\% & \RefPairedCutSevenFiveNaturalPrevalence & \RefPairedCutSevenFiveNaturalDetectorAucpr & \RefPairedCutSevenFiveNaturalEcodAucpr & \RefPairedCutSevenFiveSyntheticPrevalence & \RefPairedCutSevenFiveSyntheticDetectorAucpr & \RefPairedCutSevenFiveSyntheticEcodAucpr & \ifnum\RefPairedCutSevenFiveOrderingsAgree=1 yes\else no\fi \\
80\% & \RefPairedCutEightZeroNaturalPrevalence & \RefPairedCutEightZeroNaturalDetectorAucpr & \RefPairedCutEightZeroNaturalEcodAucpr & \RefPairedCutEightZeroSyntheticPrevalence & \RefPairedCutEightZeroSyntheticDetectorAucpr & \RefPairedCutEightZeroSyntheticEcodAucpr & \ifnum\RefPairedCutEightZeroOrderingsAgree=1 yes\else no\fi \\
85\% & \RefPairedCutEightFiveNaturalPrevalence & \RefPairedCutEightFiveNaturalDetectorAucpr & \RefPairedCutEightFiveNaturalEcodAucpr & \RefPairedCutEightFiveSyntheticPrevalence & \RefPairedCutEightFiveSyntheticDetectorAucpr & \RefPairedCutEightFiveSyntheticEcodAucpr & \ifnum\RefPairedCutEightFiveOrderingsAgree=1 yes\else no\fi \\
90\% & \RefPairedCutNineZeroNaturalPrevalence & \RefPairedCutNineZeroNaturalDetectorAucpr & \RefPairedCutNineZeroNaturalEcodAucpr & \RefPairedCutNineZeroSyntheticPrevalence & \RefPairedCutNineZeroSyntheticDetectorAucpr & \RefPairedCutNineZeroSyntheticEcodAucpr & \ifnum\RefPairedCutNineZeroOrderingsAgree=1 yes\else no\fi \\
\bottomrule
\end{tabular}
\end{table}

\subsection{ECOD's scores depend on the batch they are scored in}
\label{sec:ecodbatch}
The sweep above evaluates the same \CicidsHeldoutSize{} held-out records that
Section~\ref{sec:outcome} evaluates, under the same fitted model, and returns a
different ECOD number. The discrepancy is not noise --- ECOD is deterministic
and both figures reproduce exactly on re-execution --- and running it down
exposes a property of the reference implementation that bears on every study
that reports an ECOD number.

\noindent\begin{minipage}{\linewidth}
PyOD's \texttt{ECOD.fit} stores the training matrix, and
\texttt{decision\_function} then prepends it to whatever batch it is asked to
score before recomputing the per-column empirical CDFs
(PyOD~2.0.5, \texttt{pyod/models/ecod.py}):
\begin{verbatim}
if hasattr(self, 'X_train'):
    original_size = X.shape[0]
    X = np.concatenate((self.X_train, X), axis=0)
self.U_l = -1 * np.log(column_ecdf(X))
self.U_r = -1 * np.log(column_ecdf(-X))
skewness = np.sign(skew(X, axis=0))
...
decision_scores_ = self.O.sum(axis=1)[-original_size:]
\end{verbatim}
\end{minipage}\par\medskip

The ECDF denominator is the height of the concatenated matrix and the skewness
sign is taken over it, so a record's score is a function of which \emph{other}
records shared its \texttt{decision\_function} call. ECOD is transductive over
the scored batch, not a fitted function evaluated pointwise.

We measured the size of the effect with everything else pinned. The evaluated
index set is fixed at the same \RevEcodBatchEvalRecords{} records, the model is
fitted once on the same benign training rows, and only the records
accompanying them in the call vary, in number and therefore in composition. Scored alone, ECOD reaches
$AP=\RevEcodBatchTwoFourZeroZeroZeroZeroAucpr$; scored in the
\CicidsEcodScoringBatch-record validation-plus-test call this paper's contrast
arms use, the same records give
$AP=\RevEcodBatchFourEightZeroZeroZeroZeroAucpr$ --- a difference of
\RevEcodBatchDelta{} $AP$ on identical records with an identical model. The
full ladder has four rungs, each scoring those same
\RevEcodBatchEvalRecords{} records preceded by successively more of the
validation records immediately before the split:
$AP=\RevEcodBatchTwoFourZeroZeroZeroZeroAucpr$ at
\RevEcodBatchEvalRecords{} records,
\RevEcodBatchThreeZeroZeroZeroZeroZeroAucpr{} at \RevEcodBatchLadderOne,
\RevEcodBatchThreeSixZeroZeroZeroZeroAucpr{} at \RevEcodBatchLadderTwo, and
\RevEcodBatchFourEightZeroZeroZeroZeroAucpr{} at \CicidsEcodScoringBatch. The
value therefore rises over the first three rungs and falls at the fourth,
spanning \RevEcodBatchSpread{} $AP$: the dependence is not monotone in batch
size and cannot be corrected for by a simple adjustment.

The second of those two values is the diagnostic one: scoring the held-out
records inside the validation-plus-test batch reproduces, from an independent
pass over the archived design matrix, the value the timestamp-ordered arm of
Section~\ref{sec:outcome} reports (\SFourCicidsNaturalEcodAucpr), to the last
reported decimal. The batch composition therefore accounts for the discrepancy
completely rather than approximately, which is what distinguishes this from an
unexplained numerical difference. The threshold inherits the dependence too, because it is read off
the validation portion of the same call.

\textbf{What is and is not new here.} That ECOD mixes the training and scored
sets is known: it was reported as a defect against the reference implementation
in 2022~\citep{pyod401} and remains open, and ADBench~\citep{han2022adbench}
evaluates PyOD detectors in a fixed harness that holds the scored batch
constant, which sidesteps rather than measures the issue. The ECOD
paper~\citep{li2023ecod} is not silent on new records: its Section~IV-D states
that ECOD requires no re-training to fit them, given a relatively large sample
and no assumed data shift. What it does not state is a rule for scoring an
out-of-sample record while leaving the fitted per-dimension ECDFs unchanged
rather than recomputing them over the concatenation, and its scores are defined
for the rows of the input matrix, so the implementation is not contradicting
the specification: the specification is silent at exactly the point where the
batch enters. It is not the only unresolved question about this reference: a separate
open report~\citep{pyod561} argues that the aggregation the code applies is not
the one either the ECOD or the COPOD paper describes. We take no position on
that second report; we note only that a detector used as a fixed point of
comparison across a literature has open, unresolved questions about what it
computes. What we add is narrow and precise: a \emph{quantification} of the
effect on a real benchmark with the evaluated records held identical, and the
statement of its cross-study consequence. We searched for a published
measurement of the effect's size and found none --- the behaviour is recorded
in the implementation's issue tracker and, so far as we can establish, has been
treated as a correctness question rather than measured as an evaluation
hazard --- so we state this as a first quantification and a first statement of
the cross-study consequence, in the ordinary sense that we are aware of no
prior one. We would rather be corrected on priority than have the measurement
go unreported, and the claim is about the measurement, not about the mechanism,
which is not ours. That consequence is that published
ECOD numbers are not comparable across studies that score a different batch,
in size or composition, even on the same dataset, the same split and the same
model, and that a reported ECOD number is therefore under-specified unless the
scoring batch is reported with it. \citet{campos2016evaluation} showed for
unsupervised outlier detection generally that reported comparisons depend on
evaluation choices that papers rarely state.

We hold ourselves to that standard. Every ECOD number in this paper states the
batch it was scored in. The contrast, LITNET and sweep numbers come from a call
of the form

\begin{center}
\texttt{run\_batch\_reference('ecod',\,Xfit,\,np.vstack([Xva,\,Xte]))}
\end{center}

\noindent so for those the scoring batch is validation-plus-test:
\CicidsEcodScoringBatch{} records for both CICIDS arms
(Section~\ref{sec:outcome}), \LitnetCaptureEcodScoringBatch{} for each
per-capture LITNET stream and \LitnetPooledEcodScoringBatch{} for the pooled
composite (Section~\ref{sec:contrast-litnet}), and between
\SweepEcodScoringBatchMin{} and \SweepEcodScoringBatchMax{} across the levels
of the prevalence sweep (Section~\ref{sec:sweep}), which resamples the stream
and so changes the batch with the level. The shared-record analysis of
Section~\ref{sec:shared} reports ECOD under two batches: each arm's
\CicidsHeldoutSize-record held-out slice, restricted afterwards to the
\RevSharedRecords{} shared records, and the shared records scored as their own
\RevSharedRecords-record batch; the split sweeps above score the tail of the
stream at each cut, so their batch is the test slice itself and varies by cut.
Cross-arm comparisons in this paper hold the batch fixed in size but not in
composition: the two arms' validation-plus-test batches have the same
\CicidsEcodScoringBatch{} records in count and different records in content,
so equal size alone does not remove the batch dependence from the contrast of
Section~\ref{sec:outcome}. Holding the timestamp arm's fitted model and the
batch size fixed and changing only the batch content, by scoring the
\RefEcodCompSharedRecords{} shared records inside the timestamp-order held-out
slice and inside the day-round-robin held-out slice
(\RefEcodCompBatchRecords{} records each, differing in
\RefEcodCompDifferingRecords), moves their AUC-PR from
\RefEcodCompNaturalBatchAucpr{} to \RefEcodCompSyntheticBatchAucpr, a
difference of \RefEcodCompDeltaAucpr{} $AP$ (\RefEcodCompDeltaAucroc{}
AUC-ROC), so composition alone enters, as the mechanism implies: the empirical
CDFs are recomputed over the concatenation, whose size and content both change
every record's score. The batch effect measured on identical records,
\RevEcodBatchDelta{} $AP$ and at most \RevEcodBatchSpread{} $AP$ across the
ladder, is below the smallest reversal margin of that contrast,
\SFourCicidsNaturalEcodMinusDetectorAucpr{} $AP$ with interval lower bound
\RefBootFullNaturalMarginLo. Under its own \RevEcodBatchEvalRecords-record
batch the timestamp-order ECOD scores \RevEcodBatchTwoFourZeroZeroZeroZeroAucpr,
a margin of \SFourCicidsNaturalEcodOwnBatchMinusDetectorAucpr{} $AP$ over the
detector, so ECOD leads that arm under both batches. Cross-arm ECOD
comparisons are batch-controlled
only where the rescoring of Table~\ref{tab:shared} makes them so, by scoring
identical records as an identical batch under each arm's fitted model. The
\CicidsHeldoutSize-versus-\CicidsEcodScoringBatch{} discrepancy is visible
precisely because the split sweep and the contrast were the one pair of
analyses that did not share a batch size.

\section{The Deployed Score Ranks Worse Than Its Own Tail Term}
\label{sec:branch}

The audit of Section~\ref{sec:detector} establishes what the two branches
\emph{are}. This section measures what they \emph{do}, and the answer is a
finding in its own right rather than a corollary: the way the two branches are
combined destroys discriminative power that one of them already had.

The score is $\max(\text{tail},\,0.25\cdot P(r\le5))$, and a branch's mean
magnitude says nothing about which branch ranks. Scoring the timestamp-ordered
held-out slice with each branch alone, using the components
\texttt{update\_score} itself computed:

\begin{table}
\caption{Branch-wise discrimination on the timestamp-ordered held-out slice.
Values are recomputed from the archived per-record score dump and so differ
from Table~\ref{tab:contrast} by \RevReproNaturalDelta{} $AP$ on the deployed
composition (Section~\ref{sec:shared}).}
\label{tab:branch}
\begin{tabular}{lrr}
\toprule
scoring branch & $AP$ & AUC-ROC \\
\midrule
deployed composition & \RevBranchCombinedAucpr & \RevBranchCombinedAucroc \\
tail term only & \RevBranchTailOnlyAucpr & \RevBranchTailOnlyAucroc \\
auxiliary term only & \RevBranchAuxOnlyAucpr & \RevBranchAuxOnlyAucroc \\
\bottomrule
\end{tabular}
\end{table}

\begin{figure}
\centering
\includegraphics[width=\textwidth]{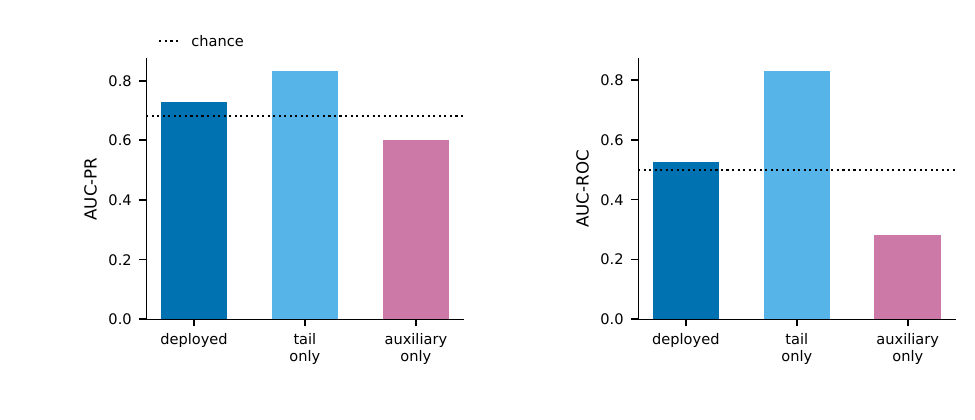}
\caption{Branch-wise discrimination on the timestamp-ordered held-out slice,
the values of Table~\ref{tab:branch}. Left: AUC-PR with the chance level at
the slice prevalence, \SFourCicidsNaturalChanceFloor. Right: AUC-ROC with the
chance level of an uninformative ranking. The deployed composition is the
detector of the other figures. The tail term alone ranks above the deployed
composition on both metrics, and the auxiliary term ranks below chance on
AUC-ROC.}
\label{fig:branch}
\end{figure}

The tail term alone \emph{outranks} the deployed composition, by
\RevBranchTailMinusCombinedAucpr{} $AP$ and
\RevBranchTailMinusCombinedAucroc{} AUC-ROC. Deleting the auxiliary branch
would therefore improve the detector on this slice. The event-block bootstrap
intervals of the three rows of Table~\ref{tab:branch} are
$[\RefBootBranchCombinedAucprLo,\,\RefBootBranchCombinedAucprHi]$ $AP$ and
$[\RefBootBranchCombinedAucrocLo,\,\RefBootBranchCombinedAucrocHi]$ AUC-ROC for
the deployed composition,
$[\RefBootBranchTailOnlyAucprLo,\,\RefBootBranchTailOnlyAucprHi]$ and
$[\RefBootBranchTailOnlyAucrocLo,\,\RefBootBranchTailOnlyAucrocHi]$ for the
tail term alone, and
$[\RefBootBranchAuxOnlyAucprLo,\,\RefBootBranchAuxOnlyAucprHi]$ and
$[\RefBootBranchAuxOnlyAucrocLo,\,\RefBootBranchAuxOnlyAucrocHi]$ for the
auxiliary term alone; the tail-only and deployed intervals do not overlap on
either metric. Figure~\ref{fig:branch}
draws the three branches against chance.

The mechanism is identified on this slice, and at the block lengths at which
the branches' AUC-PR intervals separate (\RefBootBlockLength{} and
\RefBootRobustBlockLengthA{} records; Section~\ref{sec:threats}) its sampling
uncertainty over the slice is small. The auxiliary term ranks \emph{below} chance here (AUC-ROC
\RevBranchAuxOnlyAucroc{} against $0.5$, with event-block bootstrap interval
$[\RefBootBranchAuxOnlyAucrocLo,\,\RefBootBranchAuxOnlyAucrocHi]$, entirely
below $0.5$), which means it is not noise on this slice but an inverted signal:
it scores attacks lower than benign records more often than not. Because the deployed score is a
$\max$, the auxiliary value sets the record's score whenever it exceeds the
tail value --- which is \RevBranchAuxBindsHeldoutPct\%\ of this held-out
slice, counted from the same archived components --- and the record's rank
then follows from all scores together, so an inverted signal is written into
the scores of a working one. That is why the
deployed composition's AUC-ROC (\RevBranchCombinedAucroc) sits near chance
while its tail component reaches \RevBranchTailOnlyAucroc.

Three consequences follow, and we state them at their proper strength. First,
the near-chance ranking of the deployed detector on this slice is a property of
the \emph{composition}, not of either component: a defect in how two signals
are combined, not in either signal, and it would not be found by any evaluation
that scores only the assembled detector. Second, it generalises as a warning
rather than as a measurement --- any detector that combines a calibrated signal
with an uncalibrated one under a $\max$ (or any other order statistic that lets
one branch dominate) can have the worse branch silently determine its ranking,
and no downstream metric will attribute the loss. Third, it supersedes, in the
opposite direction, any claim that the tail term simply performs the
discriminative work of the deployed score: the tail term does not merely carry
the score, it is held back by it. We state the measurement as a property of one
slice of one benchmark; the branch-wise decomposition needed to find it costs
nothing, because the components are values \texttt{update\_score} already
computes.

\section{Prevalence Within One Assembled Construction}
\label{sec:sweep}

An earlier version reported a prevalence sweep as evidence about deployment
regimes. Retained here, it is a controlled experiment \emph{on the assembled
(interleaved) CICIDS2017 stream} and nothing more: prevalence is varied by
stratified resampling at five levels, three draws each. It does not isolate
prevalence either --- only the unresampled level's draws see identical records,
so which records are retained, how many of them there are, and how far apart in
the stream the retained ones sit all vary with the nominal level. Table~\ref{tab:sweep} reports $AP$ with the achieved held-out prevalence
as the chance level beside every prevalence level.

\begin{table}
\caption{AUC-PR by prevalence level on the assembled construction (mean over
three resampling draws). ``Chance level'' is the achieved held-out prevalence.
At the unresampled level all draws see the same records; at resampled levels
the three draws of every method differ by the resampling draw. Per-draw values
and normalized lift are in the archived findings.}
\label{tab:sweep}
\begin{tabular}{lrrrrrr}
\toprule
Level & Chance level & evaluated detector & HST & LODA & ECOD (batch) & LOF (batch) \\
\midrule
5\% & \STwoLFiveFloor & \STwoLFiveProposedMean & \STwoLFiveHstMean & \STwoLFiveLodaMean & \STwoLFiveEcodMean & \STwoLFiveLofMean \\
10\% & \STwoLTenFloor & \STwoLTenProposedMean & \STwoLTenHstMean & \STwoLTenLodaMean & \STwoLTenEcodMean & \STwoLTenLofMean \\
unresampled & \STwoLUnresampledFloor & \STwoLUnresampledProposedMean & \STwoLUnresampledHstMean & \STwoLUnresampledLodaMean & \STwoLUnresampledEcodMean & \STwoLUnresampledLofMean \\
40\% & \STwoLFortyFloor & \STwoLFortyProposedMean & \STwoLFortyHstMean & \STwoLFortyLodaMean & \STwoLFortyEcodMean & \STwoLFortyLofMean \\
64\% & \STwoLSixtyFourFloor & \STwoLSixtyFourProposedMean & \STwoLSixtyFourHstMean & \STwoLSixtyFourLodaMean & \STwoLSixtyFourEcodMean & \STwoLSixtyFourLofMean \\
\bottomrule
\end{tabular}

\end{table}

Two observations. The first is at the unresampled level, where all draws see
identical records and the detector, ECOD and LOF are each deterministic, so a
flat comparison is admissible here --- identical records, no seed variance to
withhold: the batch label-privileged reference LOF
attains $AP=\STwoLUnresampledLofMean$ against the detector's
$AP=\STwoLUnresampledProposedMean$ on that identical slice.
Second, the detector's additive lift $AP-p$ runs \STwoLFiveProposedLift{} at
5\%, \STwoLTenProposedLift{} at 10\%, \STwoLUnresampledProposedLift{}
unresampled, \STwoLFortyProposedLift{} at 40\%, and
\STwoLSixtyFourProposedLift{} at 64\%, and its normalized form
$(AP-p)/(1-p)$ runs \STwoLFiveProposedNormLift, \STwoLTenProposedNormLift,
\STwoLUnresampledProposedNormLift, \STwoLFortyProposedNormLift{} and
\STwoLSixtyFourProposedNormLift{} over the same five levels --- both below zero
at the highest level, meaning below what an uninformative ranking achieves. The
additive form peaks at the 10\%\ level and the normalized form at the
unresampled one, which is why both are reported. Nothing in this section transfers to the timestamp-ordered stream or
to deployment.

\section{Provenance Discipline}
\label{sec:provenance}

Every measured value in this paper is a \LaTeX{} macro generated from
archived run manifests recording git commit, configuration hash, input
SHA-256, seed, environment hash, timestamps and output paths, and a claim
ledger maps every sentence of the abstract and introduction to its generating
run. The build fails on a measurement-shaped number with no manifested value,
on a macro two runs claim with different values, and on drift between the
derived macro index and the manifests behind it. Protocol constants that
define the experiment rather than result from it (the 70/15/15 split, budget
sizes, the auxiliary weight, library versions) and values quoted from cited
work are typed, and are identified as such where they appear. What the checks
cover, and what they do not --- the correctness of code, labels,
preprocessing or interpretation, which no provenance check reaches --- is
stated in Appendix~\ref{app:provenance}. Provenance establishes that a
displayed number came from a recorded execution on recorded inputs, and
nothing beyond that. The manifests, the macro layer, the claim ledger and the
checks ship in the artifact.

\section{Relationship to Prior Versions of This Work}
\label{sec:disclosure}

Earlier versions of this manuscript (arXiv:2605.24696 v1, posted 23 May 2026,
and v2, posted 25 June 2026) were publicly posted and are superseded by this
one. The per-result-group account of what is reused, re-derived, corrected,
withdrawn or new follows; the dated version history and the list of claims
each correction invalidates are the corrected-incident log in the artifact.

One statement can be made in the body without breaking anonymity, and is
made here because it bears on how the results should be read. Some
measurements in this paper also appear in the earlier versions, reported there
under an interpretation that this paper withdraws and replaces. Which
measurements those are, which results are new, and which earlier results are
withdrawn is as follows. The pooled LITNET composite and the assembled CICIDS arm
are the same measurements as in v1 and v2, reported there under a regime
interpretation that this paper replaces with an assembly interpretation;
the timestamp-ordered CICIDS arm, the per-capture LITNET streams, the
shared-record analysis and the method-identity audit have no counterpart
in any earlier version; and results on a third dataset that appeared in
the earlier versions are withdrawn and are not relied upon anywhere in
this paper.

The corrected-incident log in the artifact records each correction with its
evidence and closure status. The count of incidents is not itself a quality
claim.

\section{Limitations and What Would Change Our Conclusions}
\label{sec:threats}

\textbf{The attribution assumption, and what the restriction does not
bound.} The assembly treatment is uncontrolled by construction: it moves
held-out membership, prevalence and order together, so no single factor is
identified by the contrast alone. Section~\ref{sec:shared} removes the
membership difference post hoc and finds the ordering reversal does not
survive, which attributes the effect to sample membership under an assumption
the archived design does not test: that history contributes no more on the
records the arms do not share than on those they do. Stated exactly, the
restriction shows that the ordering does not reverse on a sample both arms
held out and that the detector's score on that sample moves by
\RevSharedDetectorSpread{} $AP$ between the two histories; it does not bound
history's effect on comparisons between shared and unshared records, which the
full-slice $AP$ of Section~\ref{sec:outcome} also contains, and it does not
decompose the remaining factors, since the detector's prefix and ECOD's
training rows still differ between the arms. ECOD is a label-privileged
diagnostic reference, not a competitor under equal information. Determinism
removes seed variance only.

\textbf{External validity: two corpora, one split rule.}
Two benchmarks, one split rule, one corpus per benchmark.
The CICIDS mechanism depends on that dataset's scripted per-day schedule
meeting a positional split; it does not establish the direction or magnitude of
assembly effects elsewhere. The LITNET identity depends on equal budgets and a
divisible boundary. The corpus is a \CicidsWeekRetentionPct\%\ budgeted
subsample. The timestamp-ordered held-out slice is one
\CicidsNaturalHeldoutSpanMinutes-minute window --- what a positional
chronological split yields on this corpus, stated rather than engineered
around. Timestamp order is the least transformed reference available for
CICIDS2017, not a universal gold standard: multi-sensor deployments
legitimately merge asynchronous sources, and the durable recommendation is to
preserve capture and time metadata, justify the assembly, and report sensitivity
to plausible alternatives.

\textbf{Shelved analyses, and what each would change.} Each of the following
was scoped out of this paper and is stated here with the conclusion it bears
on.
\begin{itemize}
\item \emph{A crossed history-by-sample design.} A factorial or
prevalence-matched design that varies the detector's history and the evaluated
sample independently would test the attribution assumption directly. If
history contributed more on the unshared records than on the shared ones, the
attribution of the reversal to sample membership would not hold, and the
reversal would have to be reported as a joint effect of membership and
history.
\item \emph{Event-based metrics.} Flow-wise average precision is the
convention of the compared work, and the claims here are claims about what
that convention measures under assembly. Event-based precision and recall for
time series~\citep{tatbul2018precision} would test whether the reversal is
specific to flow-wise scoring; a different ordering under an event-based
metric would narrow the reversal to the flow-wise convention without
contradicting it.
\item \emph{An imputation ablation.} The $\pm\infty$-to-zero mapping of
Table~\ref{tab:protocol} touches \RefImputeCicidsAffectedRows{} of the
\RefImputeCicidsRows{} CICIDS2017 rows and no LITNET row. An ablation that
reruns both arms without it would test whether those rows move any contrast;
the counts bound what it could find.
\item \emph{Intervals for the LITNET and sweep results.} The per-record
scores behind the LITNET per-stream values, the pooled composite and the
prevalence sweep were not archived, so those results carry no interval and
have weaker inferential support than the headline contrast. Intervals would
state whether the per-stream orderings of Table~\ref{tab:litnet} and the
sweep's lift ordering are stable under resampling of the evaluated slice; the
pooling identity is arithmetic on counts and carries no sampling
uncertainty.
\item \emph{Held-out diagnostics for the alternative reset formulation.}
The diagnostics of Appendix~\ref{app:ablation} are measured on the first
\SSixDiagWindowRows{} records of the stream and the detection metrics on the
final \SSixTestRows{} held-out records, and the held-out per-record scores were
not archived. Recomputing the diagnostics on the held-out slice would turn
``consistent with'' into a shown equality or a shown difference of the two
score distributions.
\end{itemize}

\textbf{Uncertainty.} With attack runs of median/p90/max
\StreamCicidsTwoZeroOneSevenRunMedian/\StreamCicidsTwoZeroOneSevenRunPninety/\StreamCicidsTwoZeroOneSevenRunMax{}
records an IID row bootstrap would be inappropriate, so every interval this
paper reports comes from a moving-block bootstrap~\citep{kunsch1989jackknife}
over the archived per-record scores of the evaluated slice, in stream order:
block length
\RefBootBlockLength{} records, longer than the p90 run, \RefBootResamples{}
resamples, and the 2.5th and 97.5th percentiles of the resampled statistic.
Those intervals quantify sampling uncertainty over the evaluated slice only,
not over the window selection or the subsample, and a rolling-origin procedure
is not applied. Rerunning that bootstrap at block lengths
\RefBootRobustBlockLengthA{} and \RefBootRobustBlockLengthB{} records, the
second longer than the longest attack run, changes exactly one conclusion:
every margin interval stays above zero and the auxiliary-only AUC-ROC interval
stays below $0.5$ at both lengths, but at \RefBootRobustBlockLengthB{} records
the tail-only and deployed AUC-PR intervals,
$[\RefBootBTwoSixZeroZeroBranchTailOnlyAucprLo,\,\RefBootBTwoSixZeroZeroBranchTailOnlyAucprHi]$
and
$[\RefBootBTwoSixZeroZeroBranchCombinedAucprLo,\,\RefBootBTwoSixZeroZeroBranchCombinedAucprHi]$,
overlap while the AUC-ROC intervals stay disjoint
(\RefBootBTwoFiveZeroConclusionChanges{} of nine checks change at the shorter
length and \RefBootBTwoSixZeroZeroConclusionChanges{} at the longer; the rerun
is archived in the artifact). Intervals are reported for the two margins of
Section~\ref{sec:outcome}, the margins of Table~\ref{tab:shared} and the
branch values of Table~\ref{tab:branch}; the LITNET per-stream values of
Table~\ref{tab:litnet}, the per-cut values of Table~\ref{tab:pairedcuts}, the
prevalence-level means of Table~\ref{tab:sweep} and the one-stream result of
Appendix~\ref{app:ablation} carry none, because the per-record scores behind
the LITNET, sweep and Appendix~\ref{app:ablation} results were not archived and
the per-cut values are single recomputations, so those results have weaker
inferential support than the headline contrast.
Stochastic baselines appear either as single draws marked as such or as means
over draws with the three draws stated, and the per-draw values are archived.

\textbf{Construct.} We do not claim the evaluated detector is change-point
detection, we do not claim the untuned variant characterizes change-point
detection with an informed reset, and we do not claim that ordering alone causes the
outcome change.

\section{Conclusion}
\label{sec:conclusion}

On these two benchmarks, the step that assembles capture files into an
evaluation stream is an uncontrolled experimental treatment: it changes which
records are evaluated, at what prevalence, in what order, and thereby what is
measured. On CICIDS2017 the same multiset reordered yields held-out slices
sharing \CicidsHeldoutOverlapPct\%\ of their records at prevalences
\SFourCicidsPrevShiftAbsPp{} points apart, and reverses the measured ordering
of the two deterministic scorers --- a reversal that disappears once both arms
are restricted to the records they share, attributing it, under the assumption
that history contributes no more on the unshared records than on the shared
ones, to the sample the assembly selects rather than to the order it imposes.
That restriction does not bound history's effect on comparisons between shared
and unshared records, which the full-slice value also contains. On LITNET-2020,
pooling reports an operating point no capture exhibits. The practical recommendation is not that chronology
is always right; it is that assembly belongs in the experimental design and in
the provenance record --- report capture identity, mixture weights, split
membership, held-out overlap, and sensitivity to the plausible alternatives ---
so that a reported operating point can be recognized as a property of the
assembly when it is one.

\section*{Data and Artifact Availability}
The code, stream-assembly scripts with SHA-256 verification, all run manifests,
the macro layer, the claim ledger and the corrected-incident history are
archived. The repository is public at
\url{https://github.com/MichelYsf/rcbsid-paper} (branch \texttt{rebuild/honest-v1}); the artifact is archived as version 2.2.0 of the Zenodo record lineage, doi:10.5281/zenodo.22673735, which supersedes version 2.1.0 (doi:10.5281/zenodo.22638195), version 2.0.0 (doi:10.5281/zenodo.22213264) and version 1.0.0 (doi:10.5281/zenodo.20074590).

\section*{Generative AI Usage}
The authors disclose that generative AI tools were used in producing this
work. Large language models
were used, under continuous human direction, to write and refactor analysis
scripts, to draft prose subsequently verified against the archived
measurements, and to operate an adversarial review harness that audited the
authors' own claims. No generative AI tool is an author of this work, and none
produced any reported measurement: every measured value was generated by
executed code that wrote an archived run manifest in the same execution, and the provenance gate
and claim ledger exist in part to make that boundary externally auditable. The
authors take full responsibility for the entire content of this work,
including any portion drafted with such assistance.

\section*{Companion Manuscript Disclosure}
A companion manuscript from the same research programme (arXiv:2510.09619) shares parts of the underlying codebase and
predates the audit reported here. The method-identity findings of
Section~\ref{sec:detector} apply to that shared lineage, and correcting it is
in hand. It is a public preprint and is not under review at any venue, and
there is accordingly no other editor to inform and no concurrent consideration
to declare. Its relationship to the results reported here, and the
relationship to the earlier public versions of this manuscript, are stated in
Section~\ref{sec:disclosure};
the earlier public versions are arXiv:2605.24696 v1 and v2.

\bibliography{references}
\bibliographystyle{tmlr}

\appendix

\section{An Alternative Reset Formulation}
\label{app:ablation}

Section~\ref{sec:detector} showed the evaluated reset posterior is pinned
below the cap by the published recursion. An alternative formulation places a
prior-predictive term on the reset branch, so that the change point is
modelled as occurring before $x_t$ rather than after it and the reset
hypothesis is informed by $x_t$. We implemented \textbf{one untuned instance}
of that formulation --- a Normal--Inverse-Gamma prior predictive, evaluated per
feature and summed: a Student-$t$ centred at the running global mean, with
$\nu{=}2$ degrees of freedom and squared scale $2\sigma^2$, where $\sigma^2$ is
the running global variance floored at $10^{-4}$. Its hyperparameters are
$\kappa_0{=}1$, $\alpha_0{=}1$ and $\beta_0{=}\sigma^2$, which give
$\nu{=}2\alpha_0$ and squared scale $(\beta_0/\alpha_0)(1+1/\kappa_0)$; nothing
is tuned ---
and measured both variants under a 30-minute compute cap whose reductions bind
this paragraph: one stream (udp\_flood), a \SSixPrefixRows-record prefix, one
seed. On the synthetic probe the variant responds: $P(r{=}0)$ peaks at
\SSixProbeCorrectedPeak{} after a $6\sigma$ shift against
\SSixProbeOriginalPeak{} for the evaluated detector. On real data it ranks near
chance: $AP=\SSixCorrectedAucpr$ against \SSixOriginalAucpr{} (chance level
\SSixChanceFloor; normalized lift \SSixCorrectedNormLift{} against
\SSixOriginalNormLift), with AUC-ROC \SSixCorrectedAucroc.

The prefix diagnostics of Table~\ref{tab:ablation} are consistent with that
ranking, not proof of it: they are measured on the first \SSixDiagWindowRows{}
records of the stream, while $AP$ and AUC-ROC are measured on the final
\SSixTestRows{} held-out records of the \SSixPrefixRows-record prefix, a
different population, and the per-record scores of that held-out slice were
not archived, so no equality of the two score distributions is shown. Under
the variant the posterior sits on
short runs at every step (mean $P(r{\le}5)=\SSixDiagCorrectedMeanShortRunMass$),
the $0.25\cdot P(r{\le}5)$ branch saturates, \SSixDiagCorrectedFracAtCap{} of
scores sit exactly at the $0.25$ cap, and \SSixDiagCorrectedDistinctScores{}
distinct values are emitted against \SSixDiagOriginalDistinctScores. The score
is \emph{not} constant --- it takes many values and has standard deviation
\SSixDiagCorrectedScoreSd{} --- but it ranks near chance on this sample, which
is what an AUC-ROC close to $0.5$ indicates. The two variants are degenerate in
different quantities, and we state each in the quantity measured. Below the cap
the evaluated variant's reset posterior is pinned at the hazard rate and
carries no information from the current record
(Section~\ref{sec:detector}); under this alternative it is the short-run mass
that saturates, at $P(r{\le}5)=\SSixDiagCorrectedMeanShortRunMass$, and the
score sits at the $0.25$ cap on \SSixDiagCorrectedFracAtCap{} of records. The
archived run recorded $P(r{=}0)$ for this alternative only on the synthetic
probe, so we assert nothing here about how often it resets on the stream. We
draw no conclusion about what a \emph{tuned}
prior-predictive scale would achieve; that requires a training/validation
tuning protocol and more than one stream, and selecting it on test labels is
forbidden by our protocol.

\begin{table}
\caption{One untuned alternative formulation saturates the score. Diagnostics over the first
\SSixDiagWindowRows{} records of the ablation stream; detection metrics on the
\SSixPrefixRows-record prefix (held-out \SSixTestRows{} records,
\SSixTestAttacks{} attacks, chance level \SSixChanceFloor).}
\label{tab:ablation}
\begin{tabular}{lrr}
\toprule
 & evaluated detector & one untuned variant \\
\midrule
mean posterior mass $P(r{\le}5)$ & \SSixDiagOriginalMeanShortRunMass & \SSixDiagCorrectedMeanShortRunMass \\
scores exactly at the $0.25$ cap & \SSixDiagOriginalFracAtCap & \SSixDiagCorrectedFracAtCap \\
distinct score values & \SSixDiagOriginalDistinctScores & \SSixDiagCorrectedDistinctScores \\
score standard deviation & \SSixDiagOriginalScoreSd & \SSixDiagCorrectedScoreSd \\
$AP$ on the prefix & \SSixOriginalAucpr & \SSixCorrectedAucpr \\
normalized lift $(AP-p)/(1-p)$ & \SSixOriginalNormLift & \SSixCorrectedNormLift \\
AUC-ROC on the prefix & \SSixOriginalAucroc & \SSixCorrectedAucroc \\
\bottomrule
\end{tabular}
\end{table}

\section{Provenance Discipline: What the Checks Cover}
\label{app:provenance}

\textbf{What the checks cover, stated as a scope rather than a name.} The
build fails on: a measurement-shaped number with no manifested value, anywhere
in the manuscript, in any file it includes, or in any file the claim ledger
cites; a macro two runs claim with different values; drift between the derived
macro index and the manifests behind it; a cited key with no bibliography
entry; and a derived quantity that disagrees with the values printed beside it.
Files whose purpose is to record withdrawn numbers are scanned and reported
rather than failed, because forcing a withdrawn value through the macro layer
would manifest an error as a measurement, and the few genuine non-measurements
inside enforced files sit in a named allowlist with a stated reason each.

The checks do \emph{not} cover: protocol constants and values quoted from
cited work, which are typed and identified as such; prose; or --- the important
one --- whether the code, labels, preprocessing or interpretation are correct.
Provenance establishes that a displayed number came from a recorded execution
on recorded inputs, and nothing beyond that.

We state the covered set explicitly, rather than the mechanism's name, for a
reason that generalises past this paper: a check is trusted according to what
its name claims and enforces only what it inspects, and the gap between the two
is invisible precisely because the name is reassuring --- a gate reported that
every number in this manuscript traced to a manifest while reading only the
file it had itself generated from those manifests, and a compile check reported
zero undefined references in a document carrying citations that resolved to
nothing. Both were true statements about a narrower scope than their names
implied. A reader should ask of any such check what set it actually reads.

\end{document}